# Harnessing Quantum Entanglement: Comprehensive Strategies for Enhanced Communication and Beyond in Quantum Networks

AMIT KUMAR BHUYAN, Michigan State University, bhuyanam@msu.edu
HRISHIKESH DUTTA, Michigan State University, duttahr1@msu.edu

Quantum communication represents a revolutionary advancement over classical information theory, which leverages unique quantum mechanics properties like entanglement to achieve unprecedented capabilities in secure and efficient information transmission. Unlike bits in classical communication, quantum communication utilizes qubits in superposition states, allowing for novel information storage and processing. Entanglement, a key quantum phenomenon, enables advanced protocols with enhanced security and processing power. This paper provides a comprehensive overview of quantum communication, emphasizing the role of entanglement in theoretical foundations, practical protocols, experimental progress, and security implications. It contrasts quantum communication's potential applications with classical networks, identifying areas where entanglement offers significant advantages. The paper explores the fundamentals of quantum mechanics in communication, the physical realization of quantum information, and the formation of secure quantum networks through entanglement-based strategies like Quantum Key Distribution (QKD) and teleportation. It addresses the challenges of long-distance quantum communication, the role of quantum repeaters in scaling networks, and the conceptualization of interconnected quantum networks. Additionally, it discusses strides towards the Quantum Internet, Quantum Error-Correcting codes, and quantum cryptography's role in ensuring secure communication. By highlighting the role of entanglement, this paper aims to inspire further research and innovation in secure and efficient information exchange within quantum networks.

## 1 INTRODUCTION

Classical Information Theory has been the basis for current communication which states that information is exchanged in form of bits. Information in any form, may it be text, audio or video, can be encoded as bit-strings for communication. Shannon provided the basis for physical resource required to transmit information in form of bit-strings [1], [2], [3]. Consequently, the binary nature of bits has worked for many years and all the existing technology till dates is based on it. However, with the proliferation of need for fast and secure information, classical communication is limited by the current resource availability.

Quantum communication has been extensively discussed and explored in recent literature for its enhanced processing capabilities which can facilitate improved communication capabilities. Contrary to the classical idea of binary bits, the fundamental unit of such systems are called quantum bits or *qubits* which is an analog of bits in classical communication. Qubits' ability to be in a superposition state of classical bits gives it the ability to store and process more information as compared to bits. Kimble [4] states that, "A network of quantum nodes that is linked by classical channels and comprises of $k$ nodes each with $n$ quantum bits (qubits) has a state space of dimension $k2^n$, whereas a fully quantum network has exponentially larger state space, $2^{kn}$." According to a study by Gisin [5], "The basic motivation is that quantum states code quantum information, called qubits in the case of two-dimensional Hilbert spaces, and that quantum information allows tasks to be performed that could only be achieved far less efficiently, if at all, using classical information."

With the above-stated capabilities of Quantum Communication, the question arises, “What are its practical application and where can it replace Classical Networks?” Researchers have tried to answer this question to the best of their abilities from various viewpoints viz. secure communication, channel resource maximization, transmission reliability and so on. The research on applicability of Quantum Communication has not reached to similar depth like its classical counterpart to answer this question with certainty. But the fundamental of Quantum mechanics have shown tremendous potential in tackling intrinsic problems in communication. The ability to exploit quantum entanglement [6], [7], [8], [9] and quantum superposition [7] enables quantum communication protocols [10], [11] to achieve unprecedented levels of security and information processing capabilities [10], [11], [12], [13]. By leveraging the inherent properties of qubits, such as their ability to be entangled over long distances, quantum communication allows for the secure transmission of information that is impervious to eavesdropping and tampering [14], [15], [16], [17]. In recent years, significant progress has been made in the field of quantum communication, leading to the development of various protocols and technologies. These advancements have not only deepened our understanding of quantum mechanics but have also paved the way for practical applications in fields such as cryptography, quantum teleportation, and quantum key distribution (QKD) [18], [19], [20], [21], [22], [23].

With a future goal of realizing Quantum Internet, researchers have investigated and networks have been deployed covering hundreds of kilometers [24], [25]. Though such quantum networks are in an experimental phase with vulnerabilities at different levels, substantial strides have been made towards making such networks realizable. Like, Quantum Error-Correcting (QEC) codes have been formulated to tackle with decoherence due to environmental losses during transmission [26], [27]. QEC is of great importance to achieve successful quantum teleportation [28], [29], [30], [31], which is the process of sending the state information of qubits from one quantum node to another without physically sending the qubit. One of the highlighting features of quantum communication aims to achieve successful quantum cryptography. This leverages the mechanical principle of no-cloning of qubits which is the core idea behind Quantum key Distribution (QKD) [16], [32]. The impermeability of QKD makes it useful against malevolent third-party eavesdroppers due to the uniqueness of quantum entanglement [17], [33], as shown in Figure 1.

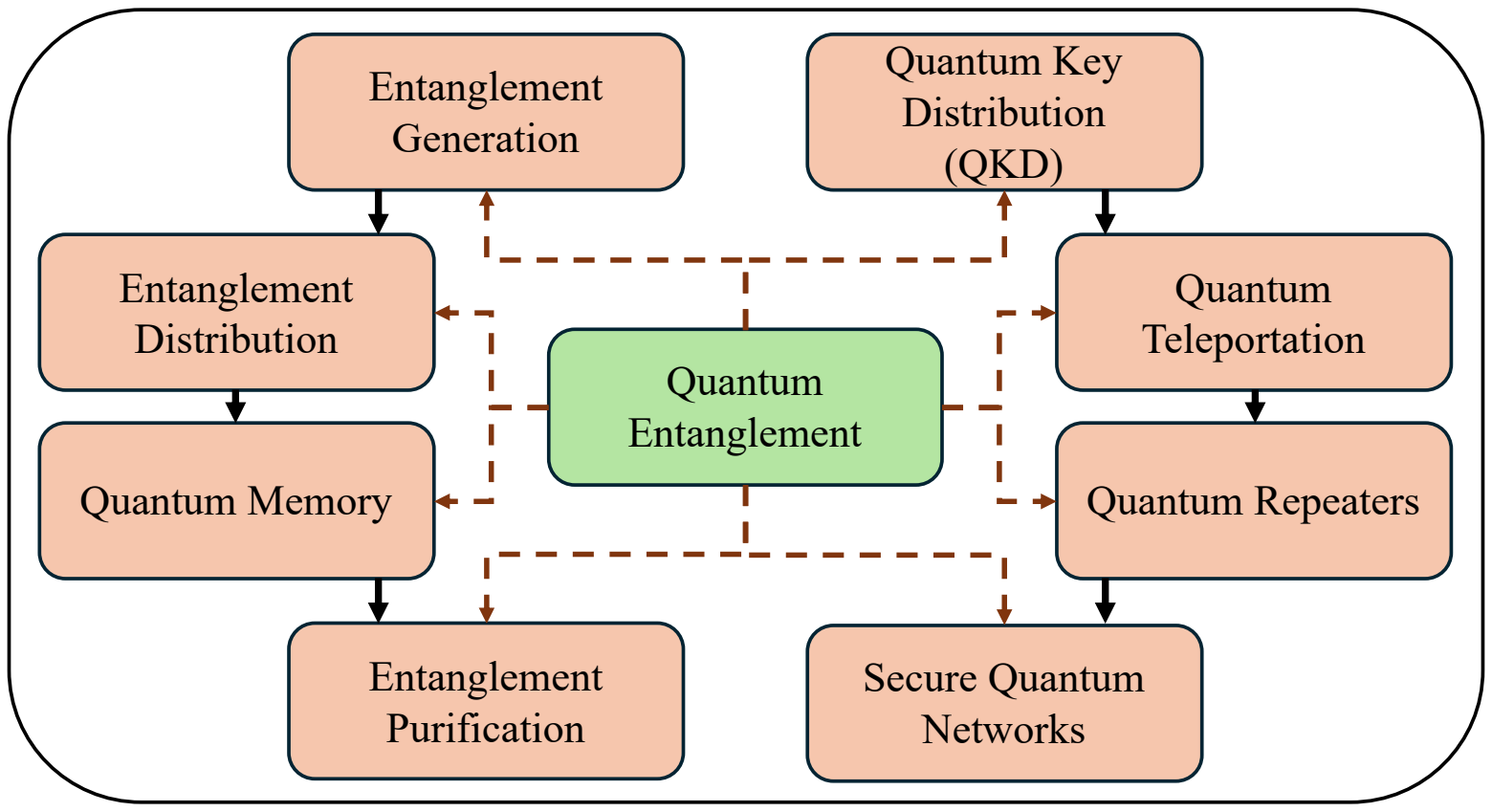

Figure 1 Quantum Entanglement at the core of successful realization a Quantum Communication Network.

Quantum communication represents a groundbreaking paradigm shift in the realm of secure information transmission. The ability to exploit the unique properties of quantum mechanics offers unparalleled advantages in terms of security, efficiency, and information processing. This review paper aims to provide a comprehensive overview of quantum communication, encompassing theoretical foundations, protocols, experimental advancements, security considerations, and future directions. By surveying the current state of the field, we hope to shed light on the remarkable achievements and challenges that lie ahead, inspiring further research and innovation in the fascinating realm of quantum communication.

In the recent years significant amount of work has been done to uncover the potential of quantum mechanics in quantum communication. There are existing surveys and reviews available that has attempted to create an ensemble of different aspects of Quantum Communication, however a more generalized and holistic review is difficult to find. The focus of the existing surveys are specific to fields like Quantum Computer and hardware [34], [35], [36], [37], [38], [39], [40], network security using Quantum Key Distribution [41], [41], [42], [43], [44], [45], [46], [47], realizing large-scale Quantum Network via entanglement and teleportation [31], [48], [49], [50], [51], [52], [53], [54] etc. The primary objective of this review paper is to provide a comprehensive overview of the state-of-the-art in quantum communication. We will delve into the fundamental concepts, protocols, and technologies that underpin this field, highlighting the key achievements and challenges encountered along the way. Additionally, we will explore the potential future directions and emerging trends in quantum communication research.

## 2 FUNDAMENTALS OF QUANTUM COMMUNICATION

This section presents the fundamentals of *Quantum Mechanics* which is the basis of a *Quantum System*. Here, we will discuss the necessary concepts to understand qubits, state vectors, quantum operators, measurements, pure and mixed states. These topics will be used to explain *Quantum Entanglement* which is an integral part of a quantum communication system.

### 2.1 A brief of Quantum Mechanics

Before getting into the mathematics of Quantum System, an understanding of Hilbert space is necessary. In quantum mechanics, the Hilbert space [55], [56], [57] refers to the mathematical space where quantum states reside. It is a complex vector space that allows us to describe and manipulate quantum systems. The specific Hilbert space associated with a quantum system depends on the number and nature of the particles involved. A vector in the Hilbert space, also known as the *state space*, represents the state of a system [58], [59].

To understand a state of a system in Hilbert space, *dirac* notation is used, also known as *bra-ket* notation [60], [61]. In *dirac* notation, the quantum states are represented as *ket* vectors $|\psi\rangle$. For example, a *ket* vector $|0\rangle$, representing the "0" state can be written as:

$$|0\rangle = \begin{pmatrix} 1 \\ 0 \end{pmatrix} \tag{1}$$

This form is behind the fundamental notion of qubit in a quantum system. To be noted that $|0\rangle$ is one of the basis states of qubit represented in a Hilbert space. For a single qubit, the Hilbert space is two-dimensional, denote as $\mathbb{C}^2$ [62], [63], because a qubit can be in a superposition of two basis states, $|0\rangle$ and $|1\rangle$. Similarly, for two qubits the Hilbert space is four-dimensional, denoted as $\mathbb{C}^4$, since it accounts for all possible combinations of the qubits i.e., $|00\rangle$, $|01\rangle$, $|10\rangle$ and $|11\rangle$. In general, for a system composed of $N$ qubits, the Hilbert space dimension is $2^N$ [64], [65], [66]. The concepts of Hilbert space and *dirac* notation are not limited to the above-mentioned details, however, this work will uncover the required concepts of Hilbert space used for Quantum communication as needed.

### 2.2 Overview of a Quantum System

As explained before, a qubit is the fundamental unit of quantum information [67]. It is the quantum analogue of a classical bit. While classical bit can be represented as a 0 or a 1, a qubit can exist in a superposition of both 0 and 1 states simultaneously. A qubit can be geometrically represented by a Block sphere, as shown in Figure 2. Any *pure* state of a qubit $|\psi\rangle$ can be represented as a point on the surface of the *Bloch sphere*. The term "*pure*" will be described in the later sections. A state in the Bloch sphere representation is determined by the polar angle $\theta$ and azimuthal angle $\varphi$, as shown in Figure 2.

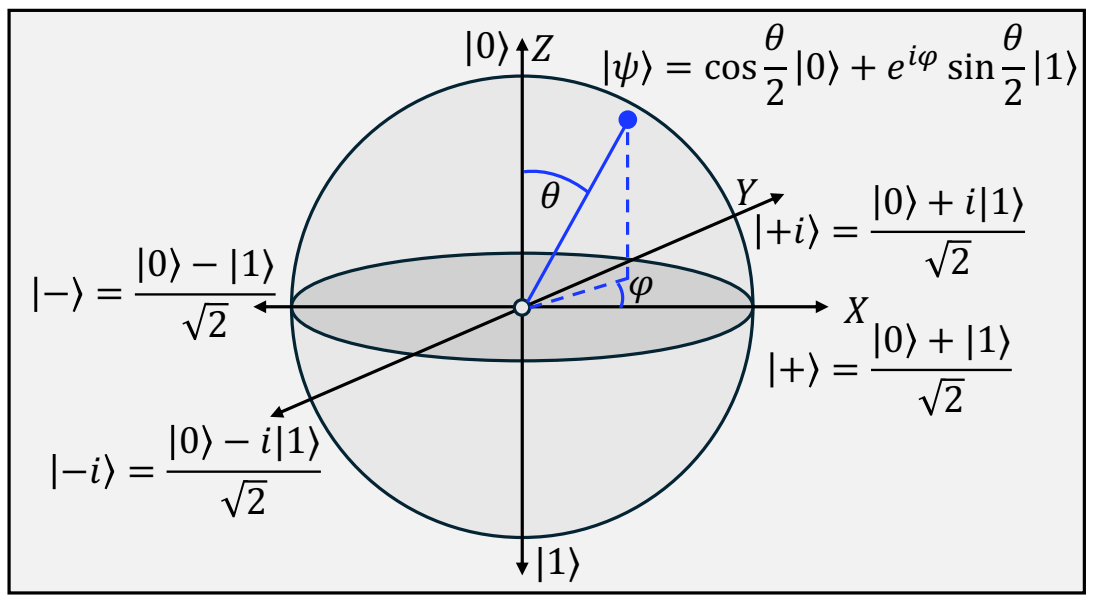

Figure 2 The Bloch sphere representation of a qubit state. The north pole is the ground state $|0\rangle$ and the south pole is the excited state $|1\rangle$. To convert an arbitrary superposition of $|0\rangle$ and $|1\rangle$ to a point on the sphere, the parametrization $|\psi\rangle = \cos\frac{\theta}{2}|0\rangle + e^{i\varphi}\sin\frac{\theta}{2}|1\rangle$ is used.

*2.2.1. Block Sphere Representation of a State.* The Bloch sphere is a sphere where each point on the surface corresponds to a unique state of the qubit. It provides an intuitive visualization of the qubit's quantum state [68], [69], [70], [71], [72]. The Bloch sphere has three axes corresponding to the three Pauli operators: X, Y, and Z [73], [74], [75]. The Z-axis represents the standard basis states $|0\rangle$ and $|1\rangle$. The state $|0\rangle$ is represented by the north pole of the sphere, and the state $|1\rangle$ is represented by the south pole [76], [77].

The X-axis is orthogonal to the Z-axis and represents the superposition of $|0\rangle$ and $|1\rangle$ [77]. The states $|+\rangle$ and $|-\rangle$ are given as follows:

$$|+\rangle = \frac{|0\rangle + |1\rangle}{\sqrt{2}} \; and \; |-\rangle = \frac{|0\rangle - |1\rangle}{\sqrt{2}} \tag{2}$$

Here $|+\rangle$ and $|-\rangle$ are represented by the points on the equator of the Bloch sphere, with $|+\rangle$ on the positive X-axis and $|-\rangle$ on the negative X-axis.

The Y-axis completes the set of orthogonal axes [76]. The states $|i\rangle$ and $|-i\rangle$ are defined as follows.

$$|i\rangle = \frac{|0\rangle + i|1\rangle}{\sqrt{2}} \; and \; |-i\rangle = \frac{|0\rangle - i|1\rangle}{\sqrt{2}} \tag{3}$$

Here $|i\rangle$ and $|-i\rangle$ are represented by the points on the equator of the Bloch sphere, with $|i\rangle$ on the positive Y-axis and $|-i\rangle$ on the negative Y-axis.

Any state of the qubit can be described by the Bloch vector, which is a point on the surface of the sphere [78], [79], [80]. The Bloch vector is characterized by two angles, $\theta$ and $\varphi$:

$$|\psi\rangle = \cos\frac{\theta}{2}|0\rangle + e^{i\varphi}\sin\frac{\theta}{2}|1\rangle \tag{4}$$

$\theta$ represents the polar angle, measured from the Z-axis (north pole) to the point on the sphere. It determines the probability amplitude of the state $|0\rangle$ and $|1\rangle$ in the qubit's superposition. $\varphi$ represents the azimuthal angle, measured around the Z-axis (longitude) to the point on the sphere. It determines the relative phase between the superposition states $|0\rangle$ and $|1\rangle$. By varying the values of $\theta$ and $\varphi$, one can represent all possible states of a qubit on the Bloch sphere [78], [79], [81]. The Bloch sphere visualization is a powerful tool for understanding and working with qubits in quantum computing.

However, a question may arise from the above explanation about states; "*Why would a qubit be in any random state*?" The answer is provided with some more complex mathematics on position of particles according to Quantum Mechanics.

*2.2.2. Uncertainty in Qubit's State.* The probabilistic nature of the state of a qubit (particle) in quantum mechanics arises from the fundamental principles of quantum theory, which are quite different from classical physics. In classical physics, the state of a particle is typically described by its position, momentum, and other observable properties, and these properties have well-defined values. If we know the initial state of a classical particle and the forces acting on it, we can predict its future state with certainty using deterministic equations of motion.

However, in quantum mechanics, particles are described by wave function, which is a complex-valued function denoted by $\psi$ [82], [83], [84]. The wave function contains all the information about the particle's quantum state, including its position, momentum, and other observable properties. However, unlike classical states, the wave function does not represent a definite value for these properties. Instead, the wave function gives the probability amplitude for the particle to have a particular outcome when measured [84], [85], [86]. The probability of finding the particle in a specific state or at a particular position is given by the squared absolute value of the wave function, $|\psi|^2$ [87]. This squared magnitude, also known as the probability density, represents the likelihood of finding the particle at different points in space. *But why is the outcome probabilistic?*

The wave function is governed by the Schrodinger equation, a fundamental equation in quantum mechanics[86], [88], [89]. The Schrodinger equation describes how the wave function evolves in time in the presence of a Hamiltonian operator, which represents the total energy of the quantum system [90], [91], [92]. Its time-dependent form is described by the following expression:

$$i\hbar\frac{\partial\psi(x,t)}{\partial t} = \mathbb{H}\psi(x,t) \Rightarrow i\hbar\frac{\partial\psi(x,t)}{\partial t} = -\frac{\hbar^2}{2m}\frac{\partial^2\psi(x,t)}{\partial x^2} + V(x)\psi(x,t) \tag{5}$$

where:

- $\psi(x,t)$ is the wave function of the particle, which depends on the position $x$ and time $t$.
- $\hbar$ is the reduced Planck constant, approximately equal to $\frac{h}{2\pi}$ (where $h$ is the Planck constant).
- $\mathbb{H}$ is the Hamiltonian operator, which acts on the wave function $\psi(x,t)$.
- $\frac{\partial}{\partial t}$ represents the partial derivative with respect to time $t$.
- $\frac{\partial^2}{\partial x^2}$ represents the second partial derivative with respect to position $x$.
- $m$ is the mass of the particle.
- $V(x)$ is the potential energy function, which depends on the position $x$.

The Schrodinger equation is a partial differential equation that relates the time derivative of the wave function to the Hamiltonian operator acting on the wave function [93], [94], [95]. The Hamiltonian operator represents the total energy of the quantum system, including its kinetic and potential energies [96], [97], [98]. In short, Schrodinger equation will help to provide information about the position of a particle from the wave function.

The probabilistic nature of quantum states means that even if a particle's initial state is known precisely, the outcome of a future measurement is uncertain [99], [100], [101], [102]. When a measurement is made, the wave function "*collapses*" into one of the possible outcomes, and the particle is found in a specific state with a probability determined by the squared absolute value of the corresponding wave function.

This intrinsic uncertainty is a fundamental aspect of quantum mechanics and is a departure from classical physics. It is a consequence of the superposition principle, where quantum systems can exist in multiple states simultaneously, and the measurement postulate, which states that the outcome of a measurement is inherently probabilistic. The probabilistic nature of quantum states is a key feature that gives rise to the rich and counterintuitive behavior observed at the quantum level.

*2.2.3. Single Qubit and Superposition.* The wave function, represented in *dirac* notation, provides a mathematical description of the quantum state, capturing both its amplitude and phase information [103], [104] (refer Figure 2). For a single qubit, the wave function is typically written as a linear combination of the basis states $|0\rangle$ and $|1\rangle$, where $|0\rangle$ and $|1\rangle$ are the two orthogonal states representing the qubit's classical states. The general form of the wave function for a single qubit can be written as:

$$|\psi\rangle = \alpha|0\rangle + \beta|1\rangle \tag{6}$$

where:

- $\psi$ is the wave function of the qubit.
- $\alpha$ and $\beta$ are complex numbers known as probability amplitudes.
- $|0\rangle$ and $|1\rangle$ are the basis states representing the qubit's classical states.

The coefficients $\alpha$ and $\beta$ determine the probability of finding the qubit in the corresponding basis states upon measurement [105]. The probability of finding the qubit in state $|0\rangle$ is given by $|\alpha|^2$, and the probability of finding it in state $|1\rangle$ is given by $|\beta|^2$ [78]. The total probability must add up to 1, meaning:

$$|\alpha|^2 + |\beta|^2 = 1 \tag{7}$$

The phase information in the wave function [106], [107], from equation 4, is important for capturing the interference effects in quantum systems, which give rise to phenomena such as superposition and entanglement, concepts that will be discussed in later sections. The relative phase between $\alpha$ and $\beta$ affects the interference patterns observed when multiple qubits are combined [108].

To determine the probabilities of measurement outcomes or to compute the behavior of quantum operations on the qubit, the wave function is manipulated using mathematical operations and applied to the relevant quantum operators.

*2.2.4. Multiple Qubits:* When we talk about multiple qubits, we refer to systems whose states are defined by more than one qubit. Similar to the state of a single qubit, the state of a system with multiple qubits is also defined by the wave function [109], [110]. When multiple qubits are combined, the state vector becomes more complex. In quantum computing, a system with two qubits can be mathematically represented using a four-dimensional complex vector. The state of a two-qubit system is described by a linear combination of basis states, where each basis state corresponds to a specific configuration of the two qubits [111]. For example, a system with two qubits can be described using a four-dimensional complex vector, such as:

$$|\psi\rangle = \alpha|00\rangle + \beta|01\rangle + \gamma|10\rangle + \delta|11\rangle \tag{8}$$

where $\alpha$, $\beta$, $\gamma$, and $\delta$ are complex numbers, and $|00\rangle$, $|01\rangle$, $|10\rangle$, and $|11\rangle$ represent the four possible basis states of the two-qubit system. As the number of qubits increases, the state space of the quantum system grows exponentially, leading to the potential for quantum computers to solve complex problems that are practically intractable for classical computers [112], [113], [114].

The basis states for a two-qubit system are denoted as $|00\rangle$, $|01\rangle$, $|10\rangle$, and $|11\rangle$, where the first digit represents the state of the first qubit, and the second digit represents the state of the second qubit. For example, $|01\rangle$ represents the state where the first qubit is $|0\rangle$ and the second qubit is $|1\rangle$. Keep in mind that the coefficients $\alpha$, $\beta$, $\gamma$, and $\delta$ must satisfy the normalization condition, which means the sum of the squared magnitudes of these coefficients must equal 1.

Let's use *dirac* notation to represent the tensor product [115], [116], [117] of two qubits to create a composite 2-qubit state. Let's say we have two individual qubits:

$$|a\rangle = \alpha|0\rangle + \beta|1\rangle,\ |b\rangle = \gamma|0\rangle + \delta|1\rangle \tag{9}$$

Their tensor product $|a\rangle \otimes |\mathrm{b}\rangle$ will result in a 2-qubit state:

$$|a\rangle \otimes |\mathrm{b}\rangle = (\alpha|0\rangle + \beta|1\rangle) \otimes (\gamma|0\rangle + \delta|1\rangle) \tag{10}$$

Using the distributive property of the tensor product, eqn. 10 is expanded as:

$$|a\rangle \otimes |\mathrm{b}\rangle = \alpha\gamma(|0\rangle \otimes |0\rangle) + \alpha\delta(|0\rangle \otimes |1\rangle) + \beta\gamma(|1\rangle \otimes |0\rangle) + \beta\delta(|1\rangle \otimes \delta|1\rangle) \tag{11}$$

Now, let's represent the tensor product of basis states:

$$|0\rangle \otimes |0\rangle = |00\rangle, |0\rangle \otimes |1\rangle = |01\rangle, |1\rangle \otimes |0\rangle = |10\rangle, |1\rangle \otimes |1\rangle = |11\rangle \tag{12}$$

An example showing the tensor product of two individual qubits $|0\rangle$ and $|1\rangle$ to make a 2-qubit state $|01\rangle$ is:

$$|0\rangle\otimes|1\rangle = \begin{pmatrix}1\\0\end{pmatrix} \otimes \begin{pmatrix}0\\1\end{pmatrix} = \begin{pmatrix}1.\begin{pmatrix}0\\1\end{pmatrix}\\0.\begin{pmatrix}0\\1\end{pmatrix}\end{pmatrix} = \begin{pmatrix}0\\1\\0\\0\end{pmatrix} = |01\rangle \tag{13}$$

The elements of the matrix correspond to the coefficients of the basis states $|00\rangle$, $|01\rangle$, $|10\rangle$, and $|11\rangle$ in a composite state. So, from eqn. 12, the combined 2-qubit state is:

$$|a\rangle \otimes |\mathrm{b}\rangle = \alpha\gamma|00\rangle + \alpha\delta|01\rangle + \beta\gamma|10\rangle + \beta\delta|11\rangle \tag{14}$$

The power of quantum computing lies in the ability to manipulate such multiple qubits to perform parallel computations and exploit interference effects [118], [119], [120]. Correlation of multiple qubits is explored to further the concept of quantum computers which is a crucial property that arises when qubits cannot be described independently. Such correlations enable quantum computers to perform certain tasks, such as factoring large numbers and searching unsorted databases, exponentially faster than classical computers.

*2.2.5. Unitary Operations:* Quantum unitary operations are fundamental transformations that preserve the inner product and the norm of quantum states, ensuring that the probabilities of measurement outcomes are conserved [121], [122], [123]. Both of these properties are explained below:

*1)* Inner Product Preservation: The inner product of two quantum states is a mathematical operation that yields a complex number. It's a measure of the similarity or correlation between two quantum states. When you apply a unitary operation to a quantum state, the inner product between that state and another state is unchanged [124]. Mathematically, if $|\psi\rangle$ and $|\phi\rangle$ are two quantum states, and $U$ is a unitary operator, then the inner product preservation property can be written as:

$$\langle\psi|\phi\rangle = \langle U\psi|U\phi\rangle \tag{15}$$

In simpler terms, the relationship between different quantum states remains the same even after applying a unitary operation [125].

*2)* Norm Preservation: The norm of a quantum state is a measure of its length or magnitude. It's obtained by taking the square root of the inner product of the state with itself. When you apply a unitary operation to a quantum state, the norm of that state remains unchanged [126]. Mathematically, if $|\psi\rangle$ is a quantum state and $U$ is a unitary operator, then the norm preservation property can be written as:

$$||U\psi|| = ||\psi|| \tag{16}$$

This means that the "length" of the state (its norm) is conserved under unitary transformations.

Unitary operators are represented by complex square matrices that satisfy the condition $U^{\dagger}U = I$, where $U^{\dagger}$ represents the conjugate transpose of the operator $U$, and $I$ is the identity matrix. Here are important types of quantum unitary operations:

*a)* Identity Operator ($I$): The identity operator leaves the quantum state unchanged [127]. Mathematically, it is represented as a 2x2 matrix:

$$I = \begin{pmatrix}1 & 0\\0 & 1\end{pmatrix} \tag{17}$$

*b)* Pauli Operators: The Pauli operators [81], [128] are a set of three unitary operators that play a fundamental role in quantum computing:

- Pauli-X flips the qubit's state:

$$X = \begin{pmatrix} 0 & 1 \\ 1 & 0 \end{pmatrix} \quad (18)$$

- Pauli-Y performs a combined bit and phase flip:

$$Y = \begin{pmatrix} 0 & -i \\ i & 0 \end{pmatrix} \quad (19)$$

- Pauli-Z performs a phase flip:

$$Z = \begin{pmatrix} 1 & 0 \\ 0 & -1 \end{pmatrix} \quad (20)$$

*c)* Hadamard Operator (*H*): The Hadamard gate creates superposition by evenly distributing probability amplitudes between $|0\rangle$ and $|1\rangle$ states [129].

$$H = \frac{1}{\sqrt{2}}\begin{pmatrix} 1 & 1 \\ 1 & -1 \end{pmatrix} \quad (21)$$

*d)* Adjoint (Hermitian Conjugate) of a Unitary Operation: The adjoint of a unitary operator $U$, denoted as $U^\dagger$, is obtained by taking the complex conjugate of the transpose of $U$ [127]. Adjoint operations are important in quantum mechanics, and they satisfy properties like $(U^\dagger)^\dagger = U$ and $(AB)^\dagger = B^\dagger A^\dagger$.

*e)* Rotation Operators: Rotation operators on the Bloch sphere correspond to unitary operations that perform rotations around different axes [130], [131]. A general rotation operator $R(\theta)$ can be represented using exponential of Pauli operators:

$$R_{\sigma_n}(\theta) = e^{i\theta\sigma_n} \quad (22)$$

where $\sigma_n$ is a Pauli operator ($X$, $Y$ or $Z$) and $\theta$ is the rotation angle.

*f)* Hermitian operators: A Hermitian operator is a linear operator that is equal to its Hermitian adjoint (conjugate transpose) [132]. In other words, an operator $O_H$ is Hermitian if and only if $O_H^\dagger = O$. Hermitian operators are fundamental in quantum mechanics because they have real eigenvalues and their eigenvectors corresponding to distinct eigenvalues are orthogonal. For example, the Pauli-Z operator is Hermitian:

$$Z = \begin{pmatrix} 1 & 0 \\ 0 & -1 \end{pmatrix}, Z^\dagger = \begin{pmatrix} 1 & 0 \\ 0 & -1 \end{pmatrix} \quad (23)$$

therefore, $Z$ is a Hermitian operator.

*g)* Unitary Operators in Bloch Sphere Representation: The Bloch sphere provides an intuitive geometric representation of a qubit's state. Each point on the surface of the sphere corresponds to a unique qubit state. Unitary operations on qubits can be visualized as rotations on the Bloch sphere [132], [133]. Suppose we have a unitary operation $U$ that acts on a qubit state $|\psi\rangle$. If $|\psi\rangle$ is represented on the Bloch sphere, the action of $U$ corresponds to a rotation of the Bloch vector representing $|\psi\rangle$. Let's consider a single-qubit state $|\psi\rangle = \alpha|0\rangle + \beta|1\rangle$. state Applying a unitary operation U to this state results in a new state $|\phi\rangle = U|\psi\rangle$. Mathematically, the action of the unitary operation U on the state can be represented as:

$$|\phi\rangle = U|\psi\rangle = \begin{pmatrix} u_{00} & u_{01} \\ u_{10} & u_{11} \end{pmatrix}\begin{pmatrix} \alpha \\ \beta \end{pmatrix} = \begin{pmatrix} u_{00}\alpha + u_{01}\beta \\ u_{10}\alpha + u_{11}\beta \end{pmatrix} \quad (24)$$

Each unitary operator can be visualized as a rotation around a specific axis by a certain angle. For example, applying the Pauli-X operator corresponds to a 180-degree rotation around the X-axis, and applying the Hadamard operator corresponds to a rotation around an axis that is located in the plane defined by the X-axis and the Z-axis on the Bloch sphere.

Some examples to solidify these manifestations are:

1. Identity Operator: $I|0\rangle = |0\rangle$

2. Pauli-X Operator: $X|0\rangle = \begin{pmatrix} 0 & 1 \\ 1 & 0 \end{pmatrix}\binom{1}{0} = \binom{0}{1} = |1\rangle$
3. Hadamard Operator: $H|0\rangle = \frac{1}{\sqrt{2}}\begin{pmatrix} 1 & 1 \\ 1 & -1 \end{pmatrix}\binom{0}{1} = \frac{1}{\sqrt{2}}\binom{1}{1} = \frac{1}{\sqrt{2}}(|0\rangle + |1\rangle) = |+\rangle$
4. Adjoint (Conjugate Transpose): $X^{\dagger} = X$
5. Rotation Operator: $R_Z\left(\frac{\pi}{2}\right)|0\rangle = \frac{1}{\sqrt{2}}(|0\rangle + i|1\rangle)$

*2.2.6. Quantum Measurements:* The axes in the Bloch sphere represents different measurement directions, also known as observables [134], [135]. In quantum mechanics, measurements involve extracting information about an observable property of a quantum system [136], [137]. Let's consider a quantum system described by a state vector $|\psi\rangle$ in a Hilbert space $\mathcal{H}$. An observable is associated and represented by a Hermitian operator $O_H$ in the Hilbert space. The eigenvalues of $O_H$ represent the possible measurement outcomes, and the corresponding eigenvectors represent the states in which measurements yield those outcomes.

Given an observable $O_H$, upon measurement the state $|\psi\rangle$ is projected onto one of the eigenstates $|o_i\rangle$ of $O_H$ [138]. The probability of measuring eigenvalue $o_i$ is given by *Born rule* [139], [140]:

$$P(o_i) = |\langle o_i|\psi\rangle|^2 \tag{25}$$

After measurement, the system's state collapses to the measured eigenstate $|o_i\rangle$. The projection operator $P_i$ projects a quantum state $|\psi\rangle$ onto a specific subspace spanned by a set of basis states. For an observable $O_H$ with eigen vector $|o_i\rangle$, the projection operator onto the $i^{th}$ eigenvector is given by:

$$P_i = |o_i\rangle\langle o_i| \tag{26 a}$$

According to this the probability of measuring eigenvalue $o_i$ can be modified as:

$$P(o_i) = |\langle o_i|\psi\rangle|^2 = \langle\psi|P_i|\psi\rangle \tag{26 b}$$

The expectation value of an observable $O_H$ in a quantum state $|\psi\rangle$ is the average value of the observable's measurements in that state [141]. In the *dirac* notation expectation value is computed as follows:

$$\langle O_H\rangle = \sum_i o_i P(o_i) = \sum_i o_i\langle\psi|P_i|\psi\rangle = \langle\psi|O_H|\psi\rangle \tag{27}$$

The variance of an observable $O_H$ in a quantum state $|\psi\rangle$ measures the spread or uncertainty of the measurement outcomes [142]. It is often associated with *quantum fluctuations* [143], [144], [145]. It is calculated as the expectation value of the squared deviation of $O_H$ from its average.

$$(\Delta O_H)^2 \equiv Var(O_H) = \langle(O_H - \langle O_H\rangle)^2\rangle = \langle O_H^2\rangle - \langle O_H\rangle^2 \tag{28}$$

Let's consider a qubit in the state $|\psi\rangle = \frac{1}{\sqrt{2}}(|0\rangle + |1\rangle)$, which is a balanced superposition of the $|0\rangle$ and $|1\rangle$ states. We will measure the spin along the X-axis using the Pauli-X operator $X$ from Equation 18. The Pauli-X operator $X$ has eigen vectors $|+\rangle = \frac{1}{\sqrt{2}}(|0\rangle + |1\rangle)$ and $|-\rangle = \frac{1}{\sqrt{2}}(|0\rangle - |1\rangle)$ corresponding to eigen values $+1$ and $-1$, respectively. The probabilities of measuring $+1$ and $-1$ as per Equations 25 and 26 are:

$$P(+1) = |\langle +|\psi\rangle|^2 = \frac{1}{2} \text{ and } P(-1) = |\langle -|\psi\rangle|^2 = \frac{1}{2} \tag{29}$$

The expected value when measuring in Pauli $X$ basis is:

$$\langle X\rangle = \langle\psi|X|\psi\rangle = (+1)P(+1) + (-1)P(-1) = \left(+\frac{1}{2}\right) + \left(-\frac{1}{2}\right) = 0 \tag{30}$$

The variance in measurement of the state is:

$$(\Delta X)^2 = \langle X^2\rangle - \langle X\rangle^2 = \langle\psi|X^2|\psi\rangle + 0 = (+1)^2P(+1) + (-1)^2P(-1) = \frac{1}{2} + \frac{1}{2} = 1 \tag{31}$$

In this example, measuring the spin of a qubit along the X-axis results in equal probabilities of getting $+1$ and $-1$ i.e., spin up and spin down, respectively. The expectation value indicates that, on average, the measured spin is zero, aligned with the Bloch sphere's equator. The measurement outcome aligns with neither $+1$ nor $-1$, which is consistent with the balanced superposition state. The variance being 1 reflects the maximum spread between the possible outcomes, indicating high uncertainty. Modern quantum measurement techniques have evolved to develop neural network based estimators where such uncertainty is traded for a systematic reconstruction bias, leading to a measurement outcome distribution with lower variance [146].

*2.2.7. Noise in Quantum Systems:* Before delving into the concepts you've mentioned, let's briefly cover the outer product notation. Given two vectors $|\psi\rangle$ and $|\phi\rangle$, their outer product is represented as $|\psi\rangle\langle\phi|$, resulting in a matrix. If $|\psi\rangle = \begin{pmatrix}\psi_1\\ \psi_2\end{pmatrix}$ and $|\phi\rangle = \begin{pmatrix}\phi_1\\ \phi_2\end{pmatrix}$, then $|\psi\rangle\langle\phi|$ is the matrix:

$$|\psi\rangle\langle\phi| = \begin{pmatrix}\psi_1\phi_1 & \psi_1\phi_2\\ \psi_2\phi_1 & \psi_2\phi_2\end{pmatrix} \tag{32}$$

Now, let's proceed with the main concepts:

*a)* Noise in Quantum State: Noise refers to unwanted or random disturbances that affect the integrity of a quantum system [147], [148]. In quantum mechanics, noise can lead to changes in the state of a quantum system, causing it to deviate from its intended state. Noise can arise due to interactions with the environment, measurement imprecision, or various other factors. Noise can cause errors in quantum computations and communications, limiting the reliability of quantum systems.

*b)* Noise in Quantum Operation: Noise in quantum operations refers to errors or inaccuracies introduced during the implementation of quantum gates or operations [149], [150]. These errors can result from imperfect control of quantum systems, decoherence (interaction with the environment), and other sources. Noise can lead to the deterioration of quantum information and affect the outcome of quantum algorithms and protocols.

*c)* Pure States: A pure quantum state is a state that can be described by a single wave function or single quantum state vector $|\psi\rangle$. Pure states are characterized by their determinism; a measurement of an observable yields a definite outcome with probability 1. Mathematically, a pure state is represented as:

$$|\psi\rangle = \begin{pmatrix}\psi_1\\ \psi_2\\ \psi_3\\ .\\ .\\ .\\ \psi_n\end{pmatrix} \tag{33}$$

Another representation of states which is pivotal in understanding states is the density matrix representation. The density matrix $\rho$ of a pure quantum state $|\psi\rangle$ is:

$$\rho = |\psi\rangle\langle\psi| \tag{34}$$

*d)* Mixed States: A mixed quantum state, also known as a statistical mixture, is a state that arises when a quantum system is in a probabilistic combination of multiple pure states [151]. A mixed state is described by a density matrix, which is a positive semidefinite Hermitian matrix. Mixed states reflect the uncertainty or lack of information about the exact state of the system and arise due to probabilistic mixtures of different pure states. Density matrices can represent mixed states along with pure states, allowing us to handle quantum states in a unified framework. The density matrix $\rho$ of a mixed quantum state with probabilities $p_i$ and pure states $|\psi_i\rangle$ is:

$$\rho = \sum_i p_i |\psi_i\rangle\langle\psi_i| \tag{35}$$

*e)* Measurement in Density Matrix Formalism: Given a density matrix $\rho$, the probability of measuring an observable $\mathcal{O}_H$ with eigen values $o_i$ is given by the trace of the product of $\rho$ and the projection operator $P_i$ onto the eigen space corresponding to $o_i$ [152], [153]:

$$P(o_i) = Tr(\rho P_i) \tag{36}$$

$$\text{where, } P_i = |o_i\rangle\langle o_i|$$

Consider a qubit in the state $|\psi\rangle = \alpha|0\rangle + \beta|1\rangle$. Its density matrix is:

$$\rho = |\psi\rangle\langle\psi| = \begin{pmatrix} |\alpha|^2 & \alpha\beta^* \\ \alpha^*\beta & |\beta|^2 \end{pmatrix} \tag{37}$$

To be noted that density matrix should satisfy the normalization condition given as:

$$Tr(\rho) = 1 \tag{38}$$

For a measurement of an observable $\mathcal{O}_H$ with eigen values $o_i$ and corresponding projection operators $P_i = |o_i\rangle\langle o_i|$, the measurement probabilities can be computed using Equations 36 and 38.

*2.2.8. Fidelity in Quantum Mechanics:* Fidelity is a concept in quantum mechanics that measures the similarity or closeness between two quantum states. It quantifies how well one quantum state can be transformed into another. Fidelity is a crucial metric in quantum information theory, quantum computation, and quantum communication, as it allows us to assess the efficiency of quantum operations and the accuracy of quantum information transmission [154], [155].

Given two quantum states $\rho$ and $\sigma$, the fidelity $F$ between them is defined as the squared absolute value of their inner product:

$$F(\rho, \sigma) = |\langle\rho|\sigma\rangle|^2 \tag{39}$$

Here, $\rho$ is the original quantum state, and $\sigma$ is the target state. It can be interpreted as follows. If $F = 1$, the states are identical or maximally similar [156]. This implies that the two states are indistinguishable, or equivalently, the two density matrices are the same. On the contrary, if $F = 0$, the states are orthogonal or completely different.

The generalized expression for fidelity in Equation 39 considers both pure and mixed states [157]. A special case of this can be expressed for fidelity of pure states, which is given by:

$$F(|\psi\rangle, |\phi\rangle) = |\langle\psi|\phi\rangle|^2 \tag{40}$$

Another extreme case manifests when both states are mixed and in the form of density matrices. Here, the fidelity expression involving mixed states $\rho$ and $\sigma$ modifies as follow:

$$F(\rho, \sigma) = \left(Tr\left(\sqrt{\sqrt{\rho}\sigma\sqrt{\rho}}\right)\right)^2 \tag{41}$$

For better comprehension, a simple example is provided, that shows in detail the computation of fidelity. Here, the original state is pure, and the measured state is mixed. Let the pure and mixed states be:

$$|\Phi^+\rangle = \frac{1}{\sqrt{2}}(|00\rangle + |11\rangle) \tag{42 a}$$

$$\rho = \frac{1}{2}(|00\rangle\langle 00| + |01\rangle\langle 01|) \tag{42 b}$$

Here, $|\Phi^+\rangle$ corresponds to a pure state whereas $\rho$ denotes a mixed state. The fidelity $F(\rho, |\Phi^+\rangle)$ between the original pure state and the measured mixed state can be calculated using the modified version of equation 39:

$$F(\rho, |\Phi^+\rangle) = \langle\Phi^+|\rho|\Phi^+\rangle \tag{43}$$

Substituting the states from equations 42 in 43:

$$\rho|\Phi^+\rangle = \left(\frac{1}{2}(|00\rangle\langle 00| + |01\rangle\langle 01|)\right) \times \left(\frac{1}{\sqrt{2}}(|00\rangle + |11\rangle)\right) \tag{44 a}$$

$$\Rightarrow \rho|\Phi^+\rangle = \frac{1}{2\sqrt{2}}(|00\rangle\langle 00|00\rangle + |01\rangle\langle 01|00\rangle + |00\rangle\langle 00|11\rangle + |01\rangle\langle 01|11\rangle) \Rightarrow \rho|\Phi^+\rangle = \frac{1}{2\sqrt{2}}(|00\rangle)$$

$$\langle\Phi^+|\rho|\Phi^+\rangle = \frac{1}{\sqrt{2}}(\langle 00| + \langle 11|) \times \frac{1}{2\sqrt{2}}(|00\rangle) = \frac{1}{4}(\langle 00|00\rangle + \langle 11|00\rangle) = \frac{1}{4}(1+0) = \frac{1}{4} \tag{44 b}$$

An alternate way of computing this converting the *dirac* notation to vector/matrix notation. Therefore, equations 42 and 43 can be rewritten as follows:

$$|\Phi^+\rangle = \frac{1}{\sqrt{2}}(|00\rangle + |11\rangle) = \frac{1}{\sqrt{2}}\begin{pmatrix}1\\0\\0\\1\end{pmatrix} \tag{45 a}$$

$$\rho = \frac{1}{2}(|00\rangle\langle 00| + |01\rangle\langle 01|) = \frac{1}{2}\begin{pmatrix}1&0&0&0\\0&1&0&0\\0&0&0&0\\0&0&0&0\end{pmatrix} \tag{45 b}$$

Substituting the states form equations 45 in equation 43:

$$\langle\Phi^+|\rho|\Phi^+\rangle = \frac{1}{\sqrt{2}}(1 \quad 0 \quad 0 \quad 1).\frac{1}{2}\begin{pmatrix}1&0&0&0\\0&1&0&0\\0&0&0&0\\0&0&0&0\end{pmatrix}.\frac{1}{\sqrt{2}}\begin{pmatrix}1\\0\\0\\1\end{pmatrix} \Rightarrow \langle\Phi^+|\rho|\Phi^+\rangle = \frac{1}{4} \tag{46}$$

In the above-mentioned example, we computed the fidelity between the given original pure state and measured mixed state. The resulting fidelity value quantifies how similar these two states are.

Fidelity satisfies the following axiomatic properties:

*1)* Range: Fidelity values lie between 0 and 1, where $F = 1$ indicates that the states are identical, and $F = 0$ means that the states are orthogonal or completely distinguishable.

*2)* Triangle Inequality: Fidelity obeys the triangle inequality, meaning that the fidelity of two states and the fidelity of each state with a third state cannot be less than the fidelity between the third state and the first two states:

$$F(\rho,\sigma) + F(\rho,\tau) \geq F(\sigma,\tau) \tag{47}$$

*3)* Fidelity and Orthogonality: If two states $\rho$ and $\sigma$ are orthogonal, their fidelity is $F(\rho,\sigma) = 0$.

*4)* Fidelity and Identity: If $\sigma = I$ (identity matrix), then $F(\rho, I)$ measures the purity of state $\rho$.

Physically fidelity is interpreted as measure that quantifies how well a quantum operation preserves the information content of a quantum state [158], [159]. It measures the overlap between the transformed state and the target state, providing a measure of how close the two states are in terms of their quantum properties.

Remember, the expression in Equation 14, where the probability amplitudes of multiple qubit states were derived from tensor product of single qubits in Equation 9? It was shown that the expression $|ab\rangle = \alpha\gamma|00\rangle + \alpha\delta|01\rangle + \beta\gamma|10\rangle + \beta\delta|11\rangle$ is combination of basis states where probability of determining the state of each qubit can be measured. Such states are called a *Product States* [160], [161]. A product state is exemplified in Figure 3. Conversely, there are correlated qubits where states of each qubit cannot be determined. They are called *Bell States* or *EPR pairs*. Such a state is shown in Figure 3. The importance of these states is dependent on an important concept in Quantum Mechanics called "*Quantum Entanglement*". It is explained next.

*2.2.9. Quantum Entanglement defying Classical Intuition:* Quantum entanglement is one of the most fascinating and counterintuitive phenomena in the field of quantum mechanics. It's a phenomenon where two or more particles become correlated in such a way that their individual states cannot be described independently [162], [163], [164]. Instead, their states become intrinsically connected, even when they are separated by large distances. This mysterious connection defies classical intuition and has profound implications for our understanding of the nature of quantum entanglement [165].

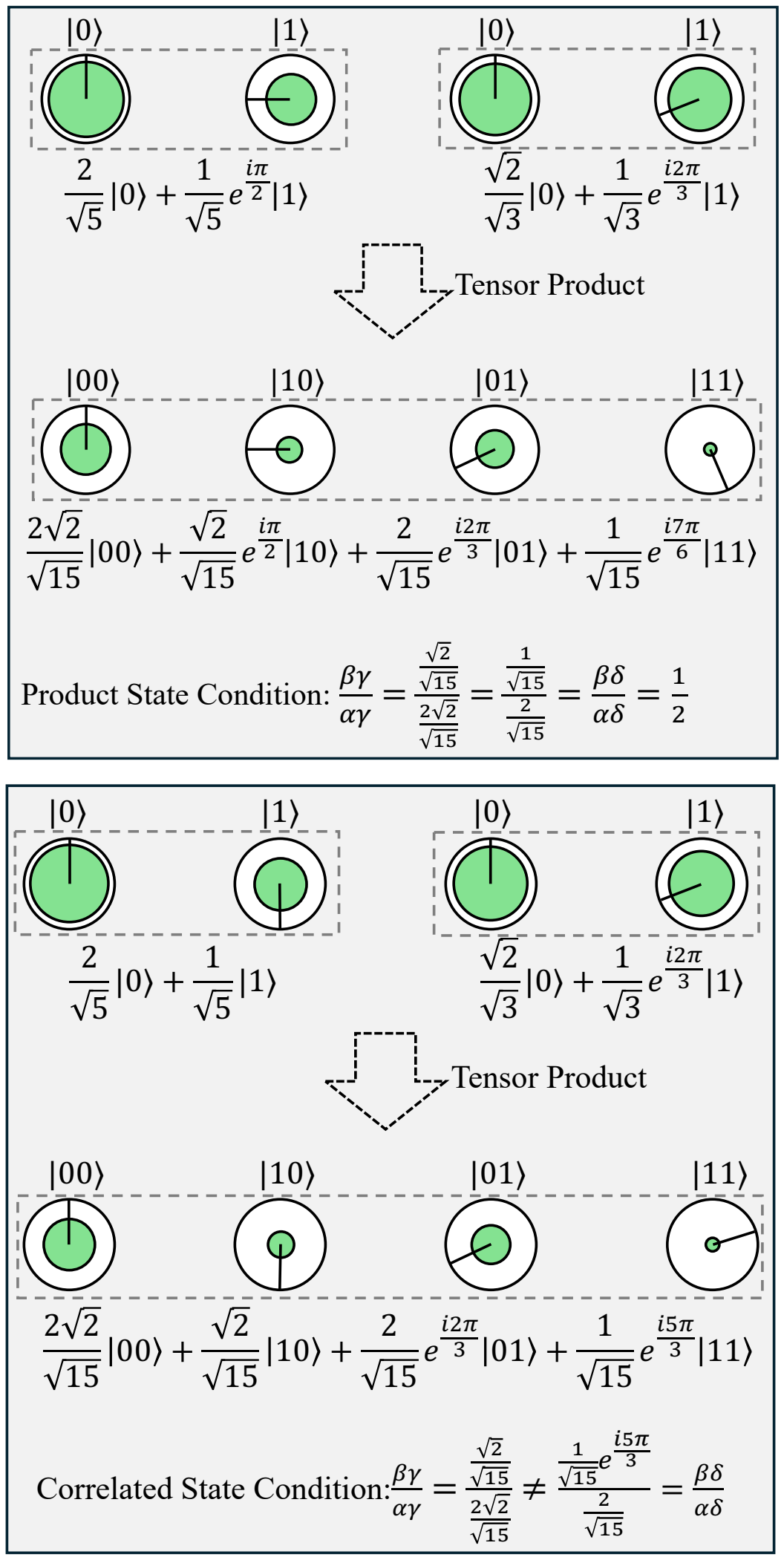

Figure 3. Product State separability (top) and Correlated State inseparability (bottom) using Dimensional Circle Notation [166]

In classical physics, objects have well-defined properties regardless of their interactions. In contrast, quantum mechanics introduces the idea that particles can exist in a superposition of multiple states until observed or measured, collapsing into a definite state upon measurement. Entanglement goes beyond superposition. When particles become entangled, measuring the state of one particle instantly affects the state of the other, regardless of the spatial separation between them [167], [168], [169]. This instantaneous connection seemingly defies the limits of information propagation imposed by the speed of light.

Entanglement is mathematically represented using the concept of a joint state, often written as a tensor product of individual states. Consider two quantum systems, represented by the states $|\psi\rangle$ an $|\phi\rangle$. Their entangled state can be written as: $|\Psi\rangle = |\psi\rangle \otimes |\phi\rangle$. This tensor product indicates that the two states are combined in a way that their individual properties are intertwined and can't be determined, leading to entanglement. The mathematical description of entanglement involves understanding how measurements on one particle provide information about the other.

To clarify the above-mentioned statement, let's consider a state:

$$|\Psi_{entangled}\rangle = \frac{1}{\sqrt{2}}(|00\rangle + |11\rangle) \tag{48}$$

Now, let's perform measurements on this entangled state. Measuring the first qubit in the standard computational basis ($|0\rangle$ or $|1\rangle$), probability of measuring $|0\rangle$ as per *Born rule* (see Equation 25):

$$P(|0\rangle) = |\langle 0|\Psi\rangle|^2 = \left|\frac{1}{\sqrt{2}}\langle 0|00\rangle + \frac{1}{\sqrt{2}}\langle 0|11\rangle\right|^2 = \left|\frac{1}{\sqrt{2}}\right|^2 = \frac{1}{2} \tag{49}$$

Similarly, probability of measuring $|1\rangle$ is given by:

$$P(|1\rangle) = |\langle 1|\Psi\rangle|^2 = \left|\frac{1}{\sqrt{2}}\langle 1|00\rangle + \frac{1}{\sqrt{2}}\langle 1|11\rangle\right|^2 = \left|\frac{1}{\sqrt{2}}\right|^2 = \frac{1}{2} \tag{50}$$

Accordingly, measuring the second qubit in the standard computational basis will give the same outcome. In this entangled state, if one qubit is measured and found it in the state $|0\rangle$, one immediately knows that the other qubit is also in the $|0\rangle$ state, and vice versa. The probabilities for each outcome are correlated.

On the other hand, individual qubits in a product state can be determined via appropriate measurement. Take the measurement of product state $|00\rangle$ in standard computational basis $|0\rangle$ and $|1\rangle$,

$$P(|0\rangle) = |\langle 0|00\rangle|^2 = 1 \tag{51}$$

$$P(|1\rangle) = |\langle 1|00\rangle|^2 = 0 \tag{52}$$

In this product state, measurements on one qubit do not provide any information about the other qubit, as shown in equations 51 and 52. The probabilities for each outcome are independent.

*2.2.10. Bell states:* Bell states are a set of four maximally entangled quantum states for a two-qubit system [170], [171]. They are named after physicist John Bell, who played a significant role in the development of quantum mechanics and in understanding the concept of entanglement [172]. The two-qubit pair is often referred to as Bell pairs or *EPR* pairs [173], [174]. The term "*EPR*" in "*EPR* pairs" refers to Einstein, Podolsky, and Rosen, who introduced the famous *EPR* paradox [175], [176], a precursor to Bell's work on entanglement. The four Bell states can be represented as follows:

$$|\Phi^+\rangle = \frac{1}{\sqrt{2}}(|00\rangle + |11\rangle) \tag{53}$$

$$|\Phi^-\rangle = \frac{1}{\sqrt{2}}(|00\rangle - |11\rangle) \tag{54}$$

$$|\Psi^+\rangle = \frac{1}{\sqrt{2}}(|01\rangle + |10\rangle) \tag{55}$$

$$|\Psi^-\rangle = \frac{1}{\sqrt{2}}(|01\rangle - |10\rangle) \tag{56}$$

With some mathematical manipulations of equations 53-56, let's express the two-qubit computational basis ($|00\rangle$, $|01\rangle$, $|10\rangle$, $|11\rangle$) in terms of Bell states ($|\Phi^+\rangle$, $|\Phi^-\rangle$, $|\Psi^+\rangle$, $|\Psi^-\rangle$):

$$|00\rangle = \frac{1}{\sqrt{2}}(|\Phi^+\rangle + |\Phi^-\rangle) \tag{57}$$

$$|11\rangle = \frac{1}{\sqrt{2}}(|\Phi^+\rangle - |\Phi^-\rangle) \tag{58}$$

$$|01\rangle = \frac{1}{\sqrt{2}}(|\Psi^+\rangle + |\Psi^-\rangle) \tag{59}$$

$$|10\rangle = \frac{1}{\sqrt{2}}(|\Psi^+\rangle - |\Psi^-\rangle) \tag{60}$$

Expressions in equation 57-60 shows that the standard computational basis states can be constructed using combinations of Bell states. It highlights the entangled nature of Bell states. The presence of Bell states in these expressions demonstrates that entanglement is fundamental to the description of quantum states.

Another important aspect of Bell states is that it gives us the ability to measure in Bell pair basis. This is especially beneficial in scenarios where qubits are entangled, and individual states can't be determined via standard computation basis, like the case shown in Equations 48-50. Equation 48 can be represented in terms of Bell pairs as follows:

$$\left|\Psi_{entangled}\right\rangle = \frac{1}{\sqrt{2}}(|00\rangle + |11\rangle) = \frac{1}{\sqrt{2}}\left(\frac{1}{\sqrt{2}}(|\Phi^+\rangle + |\Phi^-\rangle) + \frac{1}{\sqrt{2}}(|\Phi^+\rangle - |\Phi^-\rangle)\right) = |\Phi^+\rangle \qquad (61)$$

Now, measurement of the state according to *Born Rule*, can be modified by inserting Bell pair basis in place of standard computational basis [177]. Therefore, probability of measuring $|\Phi^+\rangle$ can be expressed as follows:

$$P(|\Phi^+\rangle) = \left|\left\langle\Phi^+\middle|\Psi_{entangled}\right\rangle\right|^2 \qquad (62)$$

Bell states are at the heart of quantum communication protocols [178]. They play a crucial role in various quantum information processing tasks, such as quantum teleportation [179] and quantum cryptography. Expressing computational basis states in terms of Bell states provides insights into the behavior of quantum systems during these processes which helps in designing and analyzing quantum communication schemes, including quantum key distribution.

## 3 PHYSICAL REALIZATION OF QUANTUM INFORMATION

Till now the fundamentals of Quantum Information [180] has been explained theoretically and analytically with the aid of mathematical expressions. Although these descriptions have been instrumental in citing the shift from classical theory to quantum theory, the physical manifestation of these concepts have not been unveiled.

### 3.1 Spontaneous Parametric Down-Conversion (SPDC)

This is a fascinating quantum optical phenomenon that allows the generation of entangled photon pairs. It's a nonlinear process occurring in a nonlinear crystal, such as beta barium borate (BBO) or periodically poled lithium niobate (PPLN). SPDC is a cornerstone in quantum optics and quantum information science [181], [182], [183], [184].

To initiate SPDC, a high-energy photon, known as the pump photon (P), is fired into a non-linear crystal [185]. Inside the nonlinear crystal, the pump photon (P) interacts with the crystal's nonlinear properties, leading to the creation of two low-energy photons: the signal photon (S) and the idler photon (I) [186], [187]. Within the crystal, the electrons exist in ground and excited states [188]. When the pump photon interacts with the crystal, it can excite an electron to a high energy state. In this nonlinear process, the excited electron spontaneously decays from its higher energy state to the ground state [189]. During this decay, it emits the said two photons i.e., the signal photon and the idler photon [190], [191], [192]. Energy is conserved in this process. The energy of the incoming pump photon is divided between the signal and idler photons. Their combined energy matches the energy of the pump photon. This process is described by the nonlinear susceptibility $\chi^{(2)}$ of the crystal [193], [194], [195].

According to quantum principles, the signal (S) and idler (I) photons don't have fixed polarization states but exist in superposition states, making them indeterminate until measured [196], [197], [198]. This superposition leads to the creation of entangled photon pairs [198], [199], [200]. This means that the polarization state of one photon is immediately correlated with the polarization state of the other, regardless of the physical distance separating them.

In SPDC, the quantum state of the photon pairs is described as a superposition of possible states. The pump photon (P) may have horizontal polarization ($|H\rangle$) or vertical polarization ($|V\rangle$) [201], [202], [203]. The signal photon (S) and idler photon (I) may be in various polarization states, forming superposition states. Let's use *dirac* notation for polarization states. The quantum state of the entangled photon pair can be described as:

$$|\Psi_{entangled}\rangle = \frac{1}{\sqrt{2}}(|H\rangle_S|V\rangle_I - |V\rangle_S|H\rangle_I) \quad (63)$$

where:
$|H\rangle_S$ represents a horizontally polarized signal photon.
$|V\rangle_I$ represents a vertically polarized idler photon.
$|V\rangle_S$ represents a vertically polarized signal photon.
$|H\rangle_I$ represents a horizontally polarized idler photon.
This expression shows that the signal (S) and idler (I) photons are entangled in their polarization states, and they exist in a superposition of being horizontally and vertically polarized.

The physical meaning of the above explanation is as follows. Imagine an SPDC setup that generates entangled photon pairs. If you measure the polarization of one photon (say, S), and you find it to be horizontally polarized $|H\rangle_S$, its entangled partner (idler, I) will instantaneously become vertically polarized $|V\rangle_I$ when measured, no matter how far apart they are. This phenomenon is a manifestation of quantum entanglement [197], [204], [205]. The resulting entangled states are a fundamental resource in quantum information and quantum communication, allowing for phenomena like quantum teleportation and Bell tests.

### 3.2 Entanglement as a Resource

One of the primary uses of quantum entanglement is to establish a network for quantum information transfer. Among various other applications of this phenomenon, the ability to maintain entangled state between two communicating nodes is the foundation of scaled up quantum networks.

To understand this, Clauser-Horne-Shimony-Holt (CHSH) game is used to highlight the capability of quantum entanglement [206], [207], [208]. The CHSH game is a famous quantum experiment that tests the violation of Bell inequalities and demonstrates the non-classical nature of quantum entanglement. In this game, there are two distant parties, Alice (A) and Bob (B), as well as a referee (R) who challenges them to perform measurements on entangled particles. Rules of the CHSH Game as listed below:

1. The referee randomly sends one of two bits, $x$ and $y$, to Alice and Bob, respectively. Each bit can be either 0 or 1.
2. Alice and Bob each receive a particle (qubit) and have to decide on a measurement to perform based on the bit they received.
3. After performing their measurements, Alice and Bob each send their outcomes, $a$ and $b$, to the referee.
4. The referee calculates the value $S = a \oplus b$, which is the exclusive OR (XOR) of Alice's and Bob's outcomes. If $S = x \cdot y$, Alice and Bob win; otherwise, they lose. Here, $a$ and $b$ can be either 0 or 1, and $x$, $y$, and $S$ are also binary values.

In a classical strategy, where the particles are not entangled, Alice and Bob can coordinate in such a way that they try to match their measurement outcomes with the XOR of the bits they receive, but this will lead to $S = x \cdot y$ only 75% of the time, meaning they win 75% of the games [209], [210].

On the contrary, while following a quantum strategy, Alice (A) and Bob (B) can share an entangled state, often a Bell state. Given the shared entangled state, parties 'A' and 'B' conduct standalone measurements, respectively. Let's consider an example where 'A' and 'B' share the following quantum state: $|\Phi^+\rangle = \frac{1}{\sqrt{2}}(|00\rangle + |11\rangle)$. As per the aforementioned rules to win the CHSH game, 'A' decides the measure in Z or X basis if it receives 0 or 1 respectively. 'B' measures in rotated basis $(Z + X)/\sqrt{2}$ or $(Z - X)/\sqrt{2}$ if it receives 0 or 1 respectively. The responses of 'A' and 'B' depends on the outcomes of their respective measurements. When using a quantum strategy that involves the use of entangled particles, it is possible to achieve a maximum winning rate of 85.36%. This is higher than the maximum winning rate of 75% that can be achieved using classical strategies and is a demonstration of the non-classical nature of quantum correlations. To understand this winning rate, it's important to consider the concept of the CHSH inequality, which is a mathematical expression used to quantify the correlations between measurements in the game [211], [212]. The CHSH inequality condition will be explained the Section 4.8.

The achievement of this winning rate validates the unique and non-classical properties of entangled quantum states. It is a central aspect of quantum non-locality [213], [214] and plays a significant role in the foundational experiments that test the principles of quantum mechanics. Violations of the CHSH inequality [211], [215] are experimentally observed and provide strong evidence for the quantum nature of entanglement.

## 4 FORMATION OF A SECURED QUANTUM NETWORK

Let's revisit the CHSH game from the beginning [216], [217]. We've observed that entanglement can significantly enhance the chances of players A and B winning the game. When they share an entangled state, their measurements can boost their odds of winning. However, during this measurement process, A and B inadvertently disrupt the entanglement [218] that existed between the two qubits. So, while they may secure a victory in this round, if they aim to win repeatedly, they must find a way to either restore the entanglement or create a new entangled pair that they can share in subsequent rounds of the game.

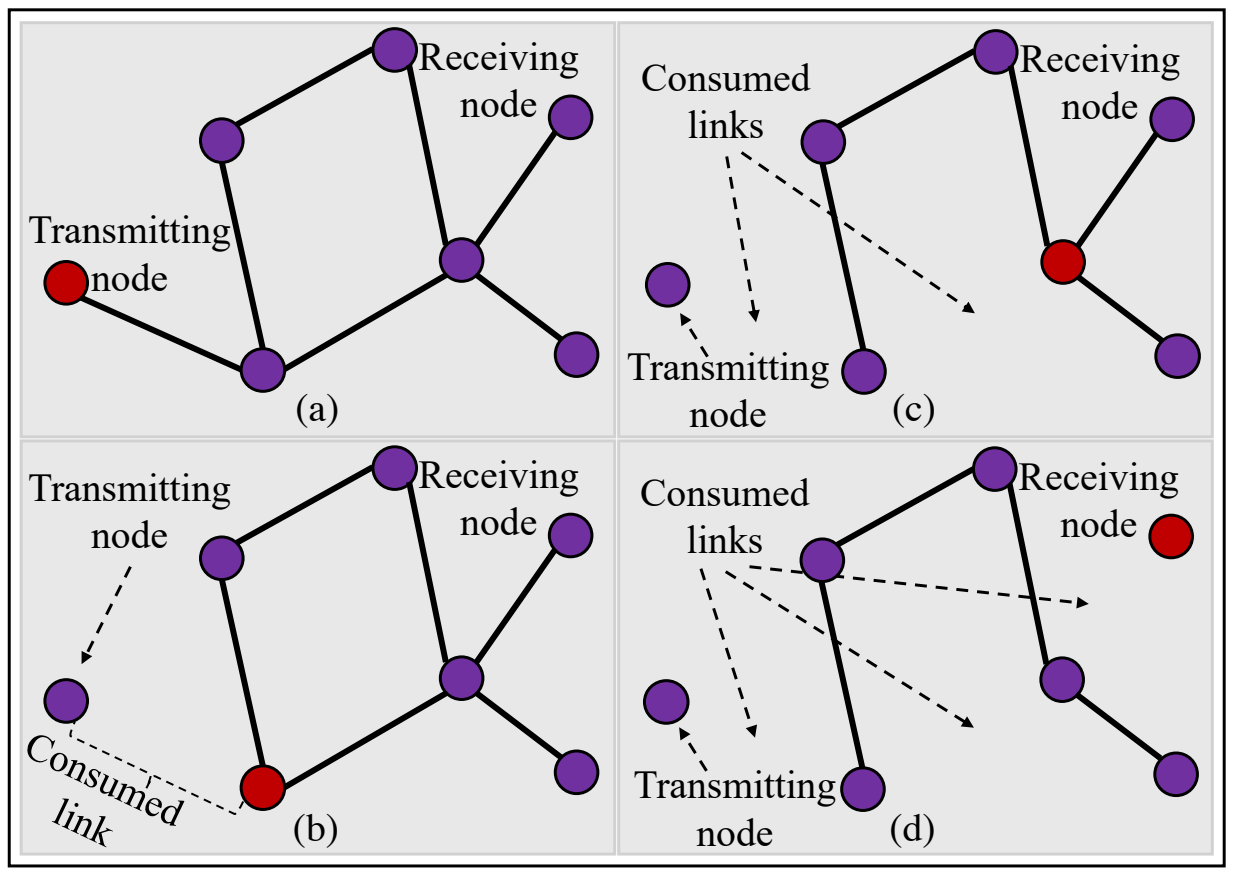

Figure 4. A Generalized Representation of Quantum Teleportation; The figure contains a basic network and the links being consumed after each measurement for the message to reach the next node. Figure 4 (a)-(d) shows the step-by-step quantum message transfer from transmitter to receiver node via Teleportation.

Similarly, entanglement plays a critical role in quantum networks. Imagine a simple quantum network, as shown in Figure 4, with red and blue dots representing network nodes, which could be qubits or collections of qubits, and black lines symbolizing the entanglement shared between these nodes. Suppose there's a sender node and a receiver node. The sender, located at one end, wishes to transmit a quantum message to the receiver at the network's opposite end.

One method to achieve this is through *teleportation* [48], [219], a process that employs measurements and entanglement to transfer information from one location to another. The sender node holds a state depicted by a red circle and aims to convey this state to the receiver node. To do so, the sender node performs a measurement, consuming the entangled link to its neighbor while transmitting the information to that node. This neighbor node repeats the same type of measurement, again consuming the entanglement between itself and its neighbor, and the process continues until the quantum information reaches the receiver node. The sender successfully delivers the message to the receiver, but the state of the network has undergone a profound transformation, as can be seen in Figure 4 (d). All the entangled links have been dissolved through the application of measurements [220]. To restore the network's full functionality, a process must be in place to reestablish the destroyed links, effectively re-establishing entanglement among them.

In this context, entanglement can be viewed as the driving force behind numerous quantum technologies. Entanglement not only enhances existing functionalities but also introduces entirely new capabilities that are absent in classical networks and classical computation. However, entanglement is a resource that can be depleted and must be restored to harness its benefits. The speed and efficiency at which we can achieve this restoration are the primary limiting factors in the development and operation of quantum networks. Before getting into the discussion about restoration of entanglement, the understanding of teleportation is necessary. The fundamentals related to the usage of quantum entanglement in teleportation and its depletion is discussed next.

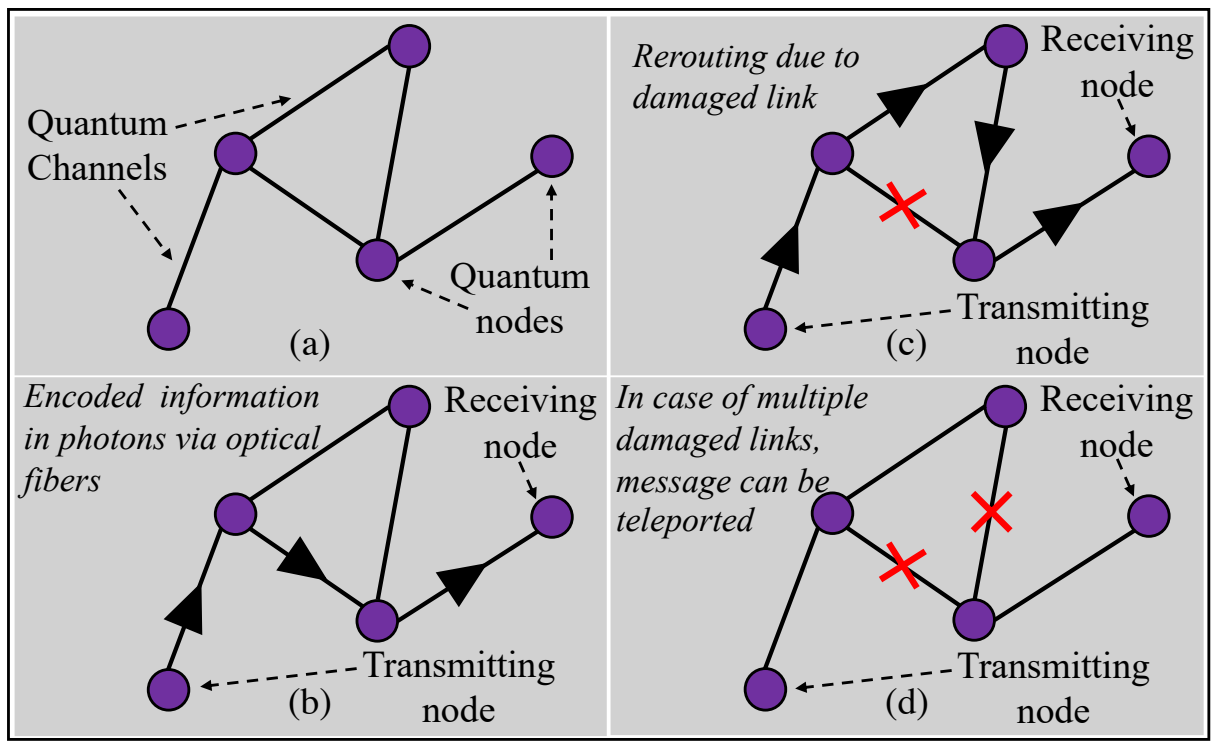

Figure 5. Need for Teleportation in case of damaged links

### 4.1 Introduction to Teleportation

Transportation stands as one of the most fundamental protocols in the field of quantum information processing. It finds extensive applications in quantum computation and quantum communication. In the context of quantum networks, a key question arises: how is the information transmitted? In classical networks, when data is downloaded or communicated over classical internet, the message essentially a bit string conveyed by photons through optical fibers. In classical networks, transmitting a message necessitates sending the physical system that encodes the message.

However, in the realm of quantum physics, the approach is somewhat different. It has the capability to transmit a message without physically transmitting the system that encodes it [221], [222]. Consider a quantum network, where the goal is to send a message. Traditionally, this would involve physically transmitting the system encoding the message. In the network diagram, as shown in Figure 5, circles represent nodes, and lines represent links between these nodes. For instance, a sender, denoted by the blue node, possesses a pure quantum state $|\psi\rangle = \alpha|0\rangle + \beta|1\rangle$ and intends to transmit this state to a target node, depicted as the orange network node. The message might be relayed through various intermediate nodes. However, if one of the quantum channels (one of the links) becomes nonfunctional, it can disrupt the message transmission. This issue can often be addressed by rerouting the message around the faulty link. Yet, what if another link fails? In this case, transmitting the message to the target node seems challenging, if not impossible, without the aid of teleportation.

Teleportation allows us to transport the state from the sender to the target without physically moving the encoding system itself [223], [224]. The physical system remains stationary while the information is transferred. An outline of the teleportation protocol [225] is as follows. It begins with sender node Alice (A) that desires to send the quantum state $|\psi\rangle$ to a node Bob (B). Both the nodes start by sharing an entangled pair of qubits. Node 'A' then performs a two-qubit measurement on its qubits, resulting in two classical bits as the measurement outcome. It communicates this outcome through a classical channel, which is possible because the result is conveyed by classical bits. Node 'B' receives this classical message, applies local corrections, and ultimately arrives at the desired state $|\psi\rangle$.

In essence, teleportation allows the transmission of quantum information while keeping the physical systems at rest [48], [226]. The next subsection contains a detailed explanation on Quantum Teleportation protocol.

### 4.2 Teleportation Protocol

Let's consider the initial state of Alice (node 'A') where it contains an arbitrary qubit as shown below:

$$|\psi\rangle = \alpha|0\rangle + \beta|1\rangle \tag{64}$$

Node 'A' also shares an entangled state (a *Bell State*) with Bob (node 'B') which is given below:

$$|\Phi^+\rangle = \frac{1}{\sqrt{2}}(|00\rangle + |11\rangle) \tag{65}$$

This state is an equal superposition of state $|00\rangle$ and $|11\rangle$. Equations 64 and 65 are diagrammatically shown in Figure 6. Given the above information, the initial state in its full form can be written as follows:

$$|\psi\rangle_{A_1}|\Phi^+\rangle_{A_2B} = (\alpha|0\rangle + \beta|1\rangle)_{A_1}\frac{1}{\sqrt{2}}(|00\rangle + |11\rangle)_{A_2B} = \frac{1}{\sqrt{2}}(\alpha|000\rangle + \alpha|011\rangle + \beta|100\rangle + \beta|111\rangle)_{A_1A_2B} \tag{66}$$

According to the teleportation protocol, Alice performs measurement in the Bell Basis. Measuring in Bell basis is equivalent to asking the question, "In which of the Bell States are Alice's qubits?". By using equations 53-56 and their manipulated form represented in equations 57-60, equation 66 can be rewritten as follows:

$$|\psi\rangle_{A_1}|\Phi^+\rangle_{A_2B} = \frac{1}{\sqrt{2}}(\alpha(|\Phi^+\rangle + |\Phi^-\rangle)|0\rangle + \alpha(|\Psi^+\rangle + |\Psi^-\rangle)|1\rangle + \beta(|\Psi^+\rangle - |\Psi^-\rangle)|0\rangle + \beta(|\Phi^+\rangle - |\Phi^-\rangle)|1\rangle)_{A_1A_2B} \tag{67}$$

If the Bell states are regrouped:

$$\begin{aligned}|\psi\rangle_{A_1}|\Phi^+\rangle_{A_2B} &= \frac{1}{2}|\Phi^+\rangle_{A_1A_2}(\alpha|0\rangle + \beta|1\rangle)_B + \frac{1}{2}|\Phi^-\rangle_{A_1A_2}(\alpha|0\rangle - \beta|1\rangle)_B \\ &= \frac{1}{2}|\Psi^+\rangle_{A_1A_2}(\alpha|1\rangle + \beta|0\rangle)_B + \frac{1}{2}|\Psi^-\rangle_{A_1A_2}(\alpha|1\rangle - \beta|0\rangle)_B\end{aligned} \tag{68}$$

Upon measurement, Alice obtains a classical result encoded as two classical bits. Importantly, the probabilities of these outcomes are independent of the probability amplitudes $\alpha$ and $\beta$. Alice communicates her measurement outcome (which Bell state she obtained) to Bob using two classical bits. These outcomes determine the state of Bob's qubit (B).

1- If Alice's measurement outcome is $\Phi^+$, then no operation is needed. Bob's qubit is already in the state $|\psi\rangle$.
2- If Alice's measurement outcome is $\Phi^-$, Bob applies a Pauli-Z gate to his qubit. Bob's qubit is in the state $Z|\psi\rangle$.
3- If Alice's measurement outcome is $\Psi^+$, Bob applies a Pauli-X gate to his qubit. Bob's qubit is in the state $X|\psi\rangle$.
4- If Alice's measurement outcome is $\Psi^-$, Bob applies the product of Pauli-X and Pauli-Z gates to his qubit. Bob's qubit is in the state $XZ|\psi\rangle$.

Now let's calculate Bob's qubit state for one of these outcomes. If Alice measures $\Phi^-$, the Bob applies Pauli-Z gate to his qubits, as shown below:

$$Z(\alpha|0\rangle + \beta|1\rangle) = \alpha|0\rangle - \beta|1\rangle \tag{69}$$

The state of Bob in Equation 69 can be verified from the measurement done in Equation 68.

These corrective operations transform Bob's qubit into the desired state $|\psi\rangle$. Teleportation thus enables the transfer of $|\psi\rangle$ from Alice to Bob without physically transmitting the quantum state. Teleportation also highlights the interchangeability of resources in quantum communication [31], [53]. It shows that by sharing an entangled pair and communicating two classical bits, Alice and Bob can accomplish the same task as sending a single qubit directly, illustrating the flexibility and power of quantum communication resources.

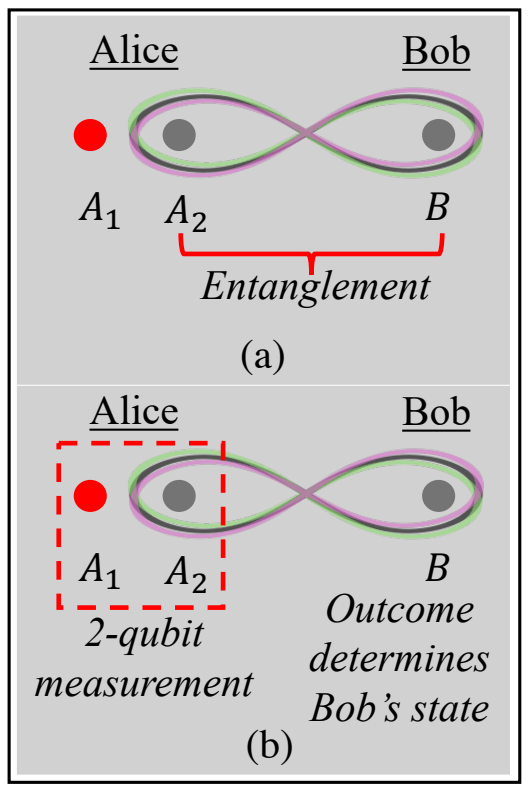

Figure 6. Teleportation Protocol: The figure shows the qubit $A_1$ at Alice that it wants to send to Bob and also shows the entangled pair $A_2B$ shared by both of them

### 4.3 No-Cloning Theorem

The No-Cloning Theorem is fundamental to understanding teleportation in quantum mechanics [227], [228]. Teleportation involves transmitting the quantum state of a particle (qubit) from one location to another without physically moving the particle itself. The No-Cloning Theorem states that an arbitrary unknown quantum state cannot be cloned exactly [229], [230]. This has profound implications for teleportation, which is explained next.

Let's revisit the teleportation protocol and address a crucial question: Did Alice clone her state while teleporting it to Bob? Even though Alice initiated with the state $|\psi\rangle$ and didn't physically transmit the qubit to Bob, at the protocol's conclusion, Bob indeed possesses the qubit $|\psi\rangle$. The key lies in the initial entanglement; Bob's qubit was part of a maximally entangled state with Alice. Post-protocol, Alice's qubit A1 becomes part of a maximally entangled state after her Bell basis measurement, projecting the two qubits onto one of the four possible Bell states, all maximally entangled.

Table 1. Comparison of Bob's State Before and After Teleportation

| | $A_1$ | $B$ |
|---|---|---|
| **Initially** | $\|\psi\rangle$ | $I/2$ |
| **After Protocol** | $I/2$ | $\|\psi\rangle$ |

This means that the state of qubit A1 initially being $|\psi\rangle$ and Bob's qubit being maximally entangled changes during the protocol, as shown in Table 1. After the protocol, Alice's qubit $A_1$ becomes maximally entangled, while Bob's qubit becomes the state $|\psi\rangle$. This transformation underscores that no cloning or copying of the state occurred, despite the absence of direct physical transmission of the state $|\psi\rangle$ from Alice to Bob.

To prove that cloning is not possible, let's delve into the question of cloning pure states [231]. Consider a hypothetical cloning device represented by a unitary operation $U$. This unitary operation acts on an arbitrary state $|\psi\rangle$ and another state initialized in state zero. The desired outcome is a two-qubit output state where both qubits are in the state $|\psi\rangle$. Some examples of the desired outcome are as follows:

$$U|0\rangle|0\rangle = |0\rangle|0\rangle \tag{70}$$

$$U|1\rangle|0\rangle = |1\rangle|1\rangle \tag{71}$$

$$U|+\rangle|0\rangle = |+\rangle|+\rangle \tag{72}$$

The question is whether such a transformation is possible, and what the unitary operator should look like. To answer this, let's consider the behavior of a *CNOT* (Controlled NOT) gate as the cloning unitary. The CNOT gate is a fundamental two-qubit gate in quantum computing. It performs a NOT operation (bit flip) on the target qubit (the second qubit) only if the control qubit (the first qubit) is in the state $|1\rangle$. If the control qubit is in the state $|0\rangle$, the target qubit remains unchanged. The matrix representation of the CNOT gate is as follows:

$$CNOT = \begin{bmatrix} 1 & 0 & 0 & 0 \\ 0 & 1 & 0 & 0 \\ 0 & 0 & 0 & 1 \\ 0 & 0 & 1 & 0 \end{bmatrix} \tag{72}$$

Now replacing CNOT in equations 70, 71 and 72, we get the following:

$$CNOT|0\rangle|0\rangle = \begin{bmatrix} 1 & 0 & 0 & 0 \\ 0 & 1 & 0 & 0 \\ 0 & 0 & 0 & 1 \\ 0 & 0 & 1 & 0 \end{bmatrix} \begin{bmatrix} 1 \\ 0 \\ 0 \\ 0 \end{bmatrix} = |0\rangle|0\rangle \tag{73}$$

$$CNOT|1\rangle|0\rangle = \begin{bmatrix} 1 & 0 & 0 & 0 \\ 0 & 1 & 0 & 0 \\ 0 & 0 & 0 & 1 \\ 0 & 0 & 1 & 0 \end{bmatrix} \begin{bmatrix} 0 \\ 1 \\ 0 \\ 0 \end{bmatrix} = |1\rangle|1\rangle \tag{74}$$

$$CNOT|+\rangle|0\rangle = \begin{bmatrix} 1 & 0 & 0 & 0 \\ 0 & 1 & 0 & 0 \\ 0 & 0 & 0 & 1 \\ 0 & 0 & 1 & 0 \end{bmatrix} \begin{bmatrix} 1 \\ 0 \\ 1 \\ 0 \end{bmatrix} = \begin{bmatrix} 1 \\ 0 \\ 0 \\ 1 \end{bmatrix} = \frac{1}{\sqrt{2}}(|00\rangle + |11\rangle) \neq |+\rangle|+\rangle \tag{75}$$

Examples in equations 70-75 demonstrate that cloning is successful for cloning $|0\rangle$ and $|1\rangle$ states but fails for superpositions of these states. To be noted that $|0\rangle$ and $|1\rangle$ are orthogonal states. To generalize this concept, let's consider arbitrary states $|\psi\rangle$ and $|\phi\rangle$. If there exists a unitary operator where cloning of states works, then the outcomes will be as follows:

$$U|\psi\rangle|0\rangle = |\psi\rangle|\psi\rangle \tag{76}$$

$$U|\phi\rangle|0\rangle = |\phi\rangle|\phi\rangle \tag{77}$$

Taking the inner products of lefthand sides and righthand sides of equations 76 and 77, we get:

$$\langle\phi|\psi\rangle = (\langle\phi|\psi\rangle)^2 \Rightarrow \langle\phi|\psi\rangle = 0 \; or \; 1 \tag{78}$$

The above mathematical analysis confirms that cloning only works for orthogonal states, showcasing a clear distinction from classical physics where arbitrary states can be cloned. This inability to clone arbitrary quantum states aligns with the No-Cloning Theorem, a fundamental principle in quantum mechanics. It asserts that an identical copy of an arbitrary unknown quantum state cannot be created. This theorem's significance comes to light in the context of teleportation.

In teleportation, entanglement between Alice and Bob is crucial. The No-Cloning Theorem ensures that this entanglement is not violated during the teleportation process. It prohibits creating an identical copy of an arbitrary unknown quantum state, which could be leveraged by eavesdroppers if cloning was possible. If perfect copies could be created, the entanglement would be compromised, leading to potential violations of quantum principles.

### 4.4 Single Photon Quantum Key Distribution

This section focuses on single-photon Quantum Key Distribution (QKD), specifically delving into the pioneering BB84 protocol. Secure communication involves three phases. Let's understand the "Three Phases of Cryptographic Secure Communication":

- *Authentication*: Initially, parties authenticate each other to confirm their identities for secure communication. Authentication follows a process where parties declare their identities and then establish trust by verifying knowledge, possession, or shared secrets.

- *Key Generation*: After authentication, the generation of a secure "key" is vital for message encryption. Different methods, like Diffie-Hellman key exchange, exist for key generation.
- *Encryption and Decryption*: The key, once generated, is used for encoding messages. Frequent key changes are essential to maintain secure encryption. Common encryption methods include Data Encryption Standard (DES), 3-DES, and Advanced Encryption Standard (AES). The one-time pad (OTP) or Vernam cipher is less common but noteworthy for its unique properties.

Encryption and decryption of messages require key generation ways that are secure. The difference between two popular categories of key generation methods, i.e., private and public keys, are given below:

- Public Key Encryption: Public key cryptography involves a public channel where Alice and Bob communicate. Alice generates a public key and a private key. Alice shares the public key with Bob, who uses it to encrypt his message. Only Alice, holding the private key, can decrypt and read the message. While functional, this method is computationally unsecure and relatively slow.
- Private Key (One-Time Pad): Alternatively, a private key generator creates a private key shared by Alice and Bob. This private key is then used for encryption and decryption. The one-time pad or Vernam cipher ensures security when the key is used only once. However, it requires a new key for each message.

Key distribution remains a challenge, especially over public channels. Quantum mechanics provides a potential solution, as explained in the subsequent steps.

### 4.5 BB84 Protocol

This protocol operates within the quantum realm, allowing Alice and Bob to utilize both a public quantum channel and a public classical channel [232], [233], [234], [235]. Here's an overview of the steps:

*1- State Preparation*:

- Alice generates two n-bit strings, denoted as "$a$" and "$b$".
- The "$b$" string dictates the basis for preparing qubits, and the "$a$" string specifies the state within that basis.
- Quantum states are created based on pairs of bits from "$a$" and "$b$", forming a composite state denoted as $|\psi\rangle$.

*2- Encoding Rules for each bit:*

- For each pair of bits from "$a$" and "$b$", Alice prepares a qubit following specific rules:
- If (0,0), she prepares the qubit in the $|0\rangle$ state, i.e., $|\psi_{00}\rangle = |0\rangle$.
- If (1,0), she prepares the qubit in the $|1\rangle$ state, i.e., $|\psi_{10}\rangle = |1\rangle$.
- If (0,1), she prepares the qubit in the $|+\rangle$ state, i.e., $|\psi_{01}\rangle = |+\rangle$.
- If (1,1), she prepares the qubit in the $|-\rangle$ state, i.e., $|\psi_{11}\rangle = |-\rangle$.
- The bit from the "$b$" string determines the basis: $Z$ basis for 0 and $X$ basis for 1.

*3- Orthogonality and Non-distinguishability:*

- Some states prepared by Alice are orthogonal, hence distinguishable. Consider states $|\psi_{00}\rangle = |0\rangle$ and $|\psi_{10}\rangle = |1\rangle$. When measured in $Z$ basis, they are distinguishable. Similarly, if the states are $|\psi_{01}\rangle = |+\rangle$ and $|\psi_{11}\rangle = |0\rangle$, they are distinguishable when measured in $X$ basis.
- Other states prepared by Alice are not orthogonal leading to non-distinguishability. Consider states $|\psi_{00}\rangle = |0\rangle$ and $|\psi_{01}\rangle = |+\rangle$. When measured in $Z$ basis, one of the outcomes can be determined with certainty and so with $X$ basis as well.
- Perfectly distinguishable states are crucial for the protocol's success.

*4- Measurement by Bob:*

- Bob receives these qubits without knowing the "$b$" string, making the states indistinguishable to him.
- He generates his random bit string "$b'$" and measures qubits in the basis determined by "$b'$".
- This generates Bob's random bit string "$a'$".

5- *Information Exchange:*
- Alice and Bob exchange information about the bases used for preparation and measurement over a public classical channel.

6- *Key Generation:*
- Bits where both Alice and Bob used the same basis, i.e., $b = b'$, are retained to form correlated bit strings "$\bar{a}$" and "$\bar{a}'$".
- These correlated bit strings serve as the shared secret key.

Table 2 shows an example implementation of the protocol. Consider a case when $n = 5$.

Table 2: Example of BB84 Implementation

| | | | | | |
|---|---|---|---|---|---|
| **String $\boldsymbol{a}$** | 1 | 0 | 0 | 1 | 1 |
| **String $\boldsymbol{b}$** | 1 | 0 | 1 | 1 | 0 |
| **Alice's Basis** | $X$ | $Z$ | $X$ | $X$ | $Z$ |
| **Encrypted qubits** | $\|-\rangle$ | $\|0\rangle$ | $\|+\rangle$ | $\|-\rangle$ | $\|1\rangle$ |
| **String $\boldsymbol{b'}$** | 1 | 1 | 1 | 0 | 0 |
| **Bob's Basis** | $X$ | $X$ | $X$ | $Z$ | $Z$ |
| **String $\boldsymbol{a'}$** | **1** | **0/1** | **0** | **0/1** | **1** |

As per the random basis determined by string "$b'$", Bob measures each qubit. The measurements where the bases of Alice and Bob match are used as a *Shared Key*. According to the above table the shared key is 101.

This protocol offers a secure key generation method, relying on the principles of quantum mechanics and correlation between the prepared and measured states. The procedure ensures the security of the generated key, paving the way for subsequent encryption and decryption steps in secure communication. Also, the protocol includes checks for security against eavesdropping, ensuring the reliability of the generated key.

### 4.6 Eavesdropper Detection using BB84 Protocol

In the preceding step, the flawless functioning of the protocol was comprehended. Now, let's delve into the impact of an eavesdropper attempting to compromise the secret key generated by Alice and Bob [236], [237], [238].

Consider the scenario where Alice communicates with Bob over a public quantum channel, and an external entity, Eve, aims to intercept their messages. The detection process is as follows:

1- *Eavesdropper's Dilemma:*
- Eve can't simply copy and resend the qubits due to the no-cloning theorem.
- To gain access, Eve must measure the qubits, but without knowing the basis information stored in Alice's "$b$".

2- *Eve's Random Basis Choice:*
- Lacking knowledge of Alice's basis information, Eve makes a random choice between $Z$ and $X$ bases for measurement.

3- *Disturbance and Detection:*
- If Eve guesses correctly (measures in the preparation basis), no disturbance occurs.
- If Eve guesses wrong, the basis changes, and disturbance happens.

*4- Detection Mechanism:*

- Alice and Bob proceed with the BB84 protocol, comparing their bases.
- Only bits measured in the same basis are retained, forming a new key.
- A portion of this key is dedicated to detecting Eve.

*5- Probability of Detection:*

- Eve has a ½ chance of measuring in the same basis as Alice.
- According to this Alice and Bob can detect Eve with a probability of ¼.

*6- Scaling with Qubit Number:*

- In an $n$ qubit system, the probability that Alice and Bob can successfully detect Eve is,

$$P(n) = 1 - \left(\frac{3}{4}\right)^n \tag{79}$$

- As the number of qubits ($n$) increases, the detection probability approaches one (refer Figure 7).

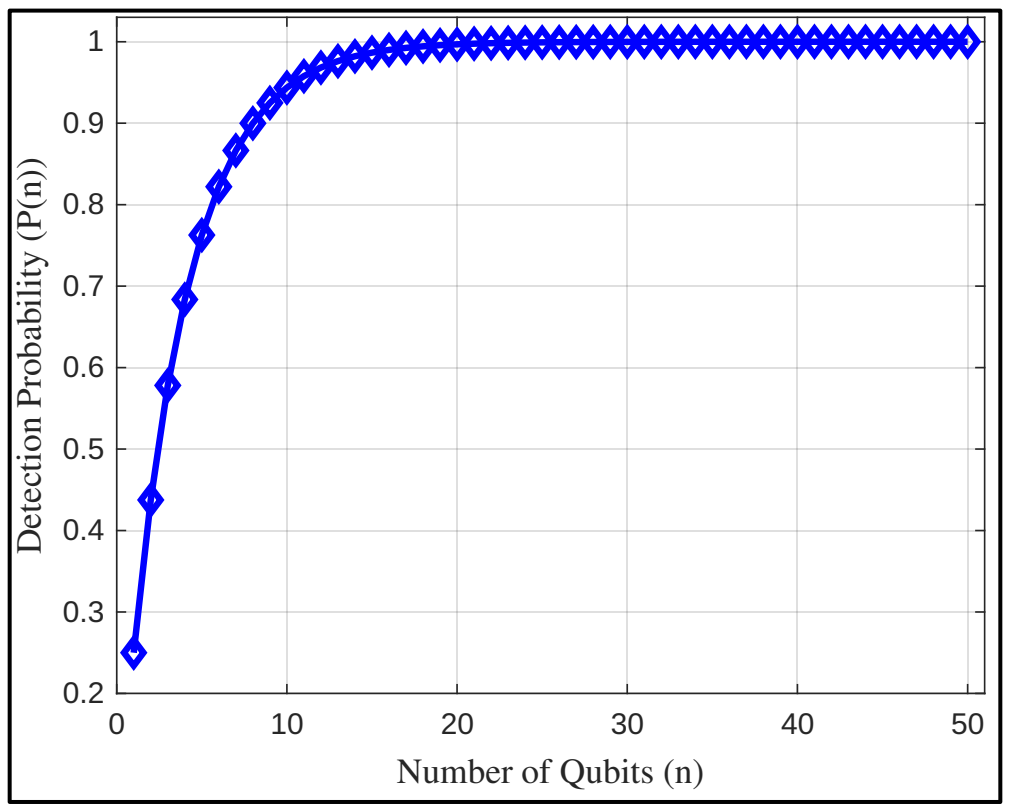

Figure 7. Probability of Detecting Eve

- Even with a modest allocation of bits (e.g., 25) for detecting Eve, the chance of undetected eavesdropping is extremely low.

*7- Decision Point:*

- With high probability, Alice and Bob can ascertain the presence of an eavesdropper.
- This detection prompts a cautious decision not to proceed with further communication, preventing unauthorized access to sensitive messages.

In essence, the protocol incorporates an effective mechanism to detect eavesdroppers. It leverages the disturbance caused when an incorrect basis is chosen, ensuring the integrity of the generated secret key.

The preceding procedure elucidated the utilization of quantum mechanics by Alice and Bob for eavesdropper detection and the generation of a secret, randomly generated key facilitating secure communication. The following experiments seek to surpass theoretical constructs, substantiating the practical realization of these concepts through the establishment and examination of authentic quantum networks.

DARPA QKD network commenced the empirical tests in 2004 at Massachusetts, USA [239]. Comprising ten nodes intricately interlinked, this network pioneered diverse communication modalities, embracing both free space and fiber optics. Unlike antecedent implementations confined to point-to-point interactions, this endeavor verified the feasibility of quantum key distribution within a network paradigm.

Subsequently, at Vienna, Europe conducted quantum network experimentation with the "Secure Communication based on Quantum Cryptography" (SECOQC) QKD network in 2008 [19]. Featuring six nodes interconnected by eight links, this network embraced a layered architectural framework, enriching the diversity of the quantum communication landscape.

A notable milestone unfolded in 2010 with the Tokyo QKD network [20], [240]. The physical manifestation of this network spanned key locations in Tokyo, including the Hongo campus at Tokyo University, Otemachi, and the NICT facility in Kogane. Distinguished by its application in a quantum secure video conference, this network exemplified the tangible application of quantum mechanics in fortifying communication within a networked environment. Such tangible quantum network testbeds have transcended theoretical frameworks, affirming the efficacy of quantum key distribution in authentic, multi-nodal scenarios within real-world networks.

### 4.7 Entanglement-based Quantum Key Distribution

Here, this paper delves into Entanglement-based Quantum Key Distribution (QKD), specifically focusing on the E91 entanglement-based protocol [41], [241].

*4.7.1. Vulnerability of BB84.* In the previous subsection, the BB84 single photon-based quantum key distribution protocol was explored. By allocating a portion of the key to eavesdropper detection, they leverage the non-orthogonality of the original qubit states. However, this protocol faces vulnerability if an eavesdropper, Eve, gains knowledge of the preparation bases, allowing her to intercept and resend qubits without detection.

*4.7.2. Modified QKD.* To enhance security, an *Entanglement-based QKD* protocol is introduced, assuming a pre-shared entanglement between Alice and Bob. In essence, communication occurs over a classical channel, and there's a source generating multiple copies of an entangled state distributed to both parties. Even with the entangled qubit source, Eve remains a potential threat. Nevertheless, the entanglement-based protocol ensures security by allowing Alice and Bob to detect Eve's eavesdropping attempts effectively. This marks a substantial improvement over the BB84 protocol, enhancing the robustness of the quantum key distribution process. The intricacies of this protocol are explored in detail.

In the architecture of Entanglement-based QKD protocol, two fundamental components come to the forefront.

1- *Establishing a Secret Key.* The initial step involves the creation of a secret key, employing an entangled state of two qubits. Let's take the scenario where Alice and Bob jointly possess a Bell pair. This state is defined by the vector $|\Psi^+\rangle = \frac{1}{\sqrt{2}}(|01\rangle + |10\rangle)_{AB}$, presenting an equal superposition of the basis states $|01\rangle$ and $|10\rangle$. When these qubits undergo measurement in the same basis, the outcomes exhibit correlation or anti-correlation, contingent upon the chosen basis. The ensuing probabilities follow a uniform random distribution. For instance, considering measurement in the $X$ basis, the probabilities of correlated outcomes (e.g., $++$ and $--$) stand at ½, and with 0 probabilities for the remaining anti-correlated outcomes scenarios.

$$Prob\{|++\rangle_{AB}\} = Prob\{|--\rangle_{AB}\} = \frac{1}{2} \tag{80}$$

$$Prob\{|+-\rangle_{AB}\} = Prob\{|-+\rangle_{AB}\} = 0 \tag{81}$$

Equations 80 and 81 shows that if Alice measures $|+\rangle_A$, then Bob always measures $|+\rangle_B$. On the other hand, if Alice measures $|-\rangle_A$, then Bob always measures $|-\rangle_B$. This process ensures the establishment of a secret random correlated key, a fusion of classical bits derived from measurement outcomes. A similar pattern unfolds when measuring in the $Z$ basis, with the distinction of anti-correlation.

$$Prob\{|01\rangle_{AB}\} = Prob\{|10\rangle_{AB}\} = \frac{1}{2} \tag{82}$$

$$Prob\{|00\rangle_{AB}\} = Prob\{|11\rangle_{AB}\} = 0 \tag{83}$$

Despite the classical key mirroring the quantum results, a simple bit-flipping operation suffices for alignment, facilitating the formation of a correlated secret random key usable for data encryption. This is shown below:

$$\begin{matrix}0_A \\ 1_B\end{matrix} \; or \; \begin{matrix}1_A \\ 0_B\end{matrix} \underset{flip\,B}{\Longrightarrow} \begin{matrix}0_A \\ 0_B\end{matrix} \; or \; \begin{matrix}1_A \\ 1_B\end{matrix} \tag{84}$$

2- *Verification of Entangled States.* The second critical element involves confirming the presence of an entangled state. This verification serves a dual purpose. Firstly, entangled states facilitate the generation of a correlated random key. Additionally, the inherent security of maximally entangled states, attributed to the "*monogamy of entanglement*", fortifies the protocol against security threats.

The monogamy of entanglement is a fundamental quantum state property constraining the correlation extent between multiple qubits [206]. If Alice and Bob share a maximally entangled state, any correlation with a third party, such as Eve, is precluded. This property becomes pivotal in demonstrating the security of the established key, ensuring Eve's lack of access to the secret key information. Striking a balance, a trade-off exists; when Alice and Bob enhance the entanglement between their particles to strengthen their communication, it has a consequence of reducing the correlations that Eve can exploit. On the other hand, if they choose to minimize correlations with Eve, it might limit the degree of entanglement between Alice and Bob, potentially impacting the strength of their quantum communication. The best-case scenario, where they become maximally entangled, assures complete absence of correlations with Eve, amplifying the security of the key shared between Alice and Bob.

### 4.8 Clauser-Horne-Shimony-Holt (CHSH) Inequality

To ascertain the maximally entangled nature of the shared key, the CHSH inequality comes into play [242], [243]. Involving a function with 4 classical random variables $A$, $\bar{A}$, $B$ and $\bar{B}$, this inequality imposes constraints on the function's average values. The maximum and minimum value this function can attain is $+2$ and $-2$ respectively, which is shown below:

$$\max |A(B + \bar{B}) + \bar{A}(B - \bar{B})| = 2 \tag{85}$$

If these random variables are constantly generated, then the average value expression, otherwise called the *CHSH* inequality expression, is given by:

$$S = |\langle A(B + \bar{B})\rangle - \langle \bar{A}(B - \bar{B})\rangle| \leq 2 \Rightarrow S = |\langle AB\rangle + \langle A\bar{B}\rangle + \langle \bar{A}B\rangle - \langle \bar{A}\bar{B}\rangle| \leq 2 \tag{86}$$

In the quantum scenario, where these variables denote measurement outcomes in a specific basis on a quantum state $\psi$, violation of the CHSH inequality, quantified by an expression denoted in Eq. 87, confirms entanglement.

$$S = |\langle AB\rangle + \langle A\bar{B}\rangle + \langle \bar{A}B\rangle - \langle \bar{A}\bar{B}\rangle| \leq 2\sqrt{2} \tag{87}$$

The maximal violation of this inequality, with $S = 2\sqrt{2}$, certifies the maximally entangled state shared by Alice and Bob. This is shown via. an example scenario as follows. Consider a maximally entangled Bell state $|\Psi^+\rangle = \frac{1}{\sqrt{2}}(|01\rangle + |10\rangle)$. The state expression and measurement expressions in their respective matrix representations are given below.

$$|\Psi^+\rangle = \frac{1}{\sqrt{2}}(|01\rangle + |10\rangle) = \begin{pmatrix}0\\1\\1\\0\end{pmatrix}$$

$$A = Z = \begin{pmatrix}1 & 0\\0 & -1\end{pmatrix};\; \bar{A} = X = \begin{pmatrix}0 & 1\\1 & 0\end{pmatrix};\; B = \frac{1}{\sqrt{2}}(Z - X) = \begin{pmatrix}1 & -1\\1 & 1\end{pmatrix};\; \bar{B} = \frac{1}{\sqrt{2}}(Z + X) = \begin{pmatrix}1 & 1\\1 & -1\end{pmatrix}$$

Now, applying the expression in equation 87 and calculating each component requires computation of expectation i.e., $\langle AB\rangle$, $\langle A\bar{B}\rangle$, $\langle \bar{A}B\rangle$ and $\langle \bar{A}\bar{B}\rangle$. For brevity, computation of one of the expectations is shown below.

$$\langle A\bar{B}\rangle = \langle \Psi^{+}|A\otimes \bar{B}|\Psi^{+}\rangle = \frac{1}{\sqrt{2}}\begin{pmatrix}0 & 1 & 1 & 0\end{pmatrix}\begin{pmatrix}1 & 1 & 0 & 0\\ 1 & -1 & 0 & 0\\ 0 & 0 & -1 & -1\\ 0 & 0 & -1 & 1\end{pmatrix}\begin{pmatrix}0\\ 1\\ 1\\ 0\end{pmatrix}$$

Similarly, computing $\langle AB\rangle$, $\langle \bar{A}B\rangle$, $\langle \bar{A}\bar{B}\rangle$ and plugging the values in equation 87 produces $S = 2\sqrt{2}$. To be noted that the CHSH inequality expression used here is exclusive to Bell State $\Psi^{+}$. To ensure the violation of CHSH inequality for other Bell Pairs different expressions are used. It is crucial to use the appropriate CHSH inequality expression to verify maximum entanglement. This verification, rooted in mathematical rigor, underlines the robustness of Entanglement-based QKD protocol.

### 4.9 E91 Quantum Key Distribution Protocol

The E91 QKD protocol amalgamates two pivotal components, previously elucidated in Section 4.7, i.e., Establishing a Secret Key and Verification of Entangled States. Let's delve into the technical integration of these components.

*4.9.1. Communication Setup.* Alice and Bob engage in communication via a classical channel. Multiple instances of a maximally entangled state are shared. These entangled states can be generated by Eve, introducing a potential security challenge.

*4.9.2. Random Measurement Bases.* Alice and Bob embark on the quantum phase by randomly selecting measurement bases for their qubits. Alice's measurement comprises of three bases i.e., $A_1$ $(Z\text{-}basis)$, $A_2$ $(X\text{-}basis)$ and $A_3$ $(Z + X\ rotated\ basis)$. Whereas Bob's measurement bases include $B_1$ $(Z\text{-}basis)$, $B_2$ $(Z - X\ rotated\ basis)$ and $B_3$ $(Z + X\ rotated\ basis)$.

*4.9.3. Key Generation.* The presence of three measurement bases for both parties is a strategic choice. Overlapping bases enable the creation of a classical correlated key when both Alice and Bob measure the entangled state in the same basis. To establish the key, Alice measures in $A_1$ or $A_3$ and Bob measures in in $B_1$ or $B_3$. Information about measurement bases is shared, and instances where bases coincide are utilized to generate a secret correlated key.

*4.9.4. CHSH Expression Computation.* Instances where measurement bases do not coincide are not discarded. Instead, they are employed to compute the CHSH expression, a critical step in verifying entanglement. Scenarios like $(A_1, B_3)$, $(A_1, B_2)$, $(A_2, B_2)$ and $(A_2, B_3)$ are sought to calculate the CHSH expression.

*4.9.5. Verification and Decision.* If the CHSH expression is less than or equal to two, certainty about the entangled state is undetermined, hence the protocol is aborted. Whereas, if the CHSH expression exceeds two, Alice and Bob confidently proceed with the protocol, capitalizing on the assurance provided by monogamy of entanglement.

In real-life scenarios, noise or other factors prevent the generation of a perfectly correlated key [244], [245]. Alice and Bob must decide on an acceptable security risk and may engage in information reconciliation and privacy amplification protocols to enhance key correlation and security.

Note that the security of the BB84 protocol is jeopardized if Eve is involved in generating the random bit string "B". In contrast, the E91 protocol's secret key is generated only after Alice and Bob measure their entangled qubits, making the key unconditionally secure. Entanglement proves to be essential for security in the E91 protocol.

### 4.10 Distributed and Blind Quantum Computation

Secured quantum communication inherently requires secured quantum computation capabilities. In the field of quantum computing, it's not necessary for a single party to conduct all quantum computations on their own, especially if they have limited resources such as an insufficient number of qubits or qubits of inadequate quality. This limitation does not prevent the execution of quantum computations. Instead, collaboration with other parties who have their own quantum computing resources can enable larger and more complex quantum computations than could be performed individually. This collaborative approach is known as "*Distributed Quantum Computation*" [246], [247], [248], [249], [250].

Consider a scenario where individuals named Alice, Bob, Charlie, Dave, and Eve each possess a small quantum computer. By pooling their resources, they can perform quantum computations that none could achieve independently. They can share and entangle qubits and exchange quantum information over a quantum network, connecting their local systems into a larger, combined system for computation. Key questions in this context include the most efficient methods for networking quantum processors, the required communication for coordinating efforts, and trust concerns among participants.

Moreover, the scenario explores situations where an individual, like Alice, may have extremely limited or no quantum computing resources but still wishes to perform quantum computations. For instance, Alice might delegate her quantum computation tasks to Bob, who possesses a powerful quantum computer. She communicates her computational requirements to Bob using classical information, which includes the input, the process, and then receives the output from Bob. However, this process reveals all details of the computation to Bob, which might not be desirable if the information is sensitive or proprietary.

To address privacy concerns, a method known as "*Blind Quantum Computation*" is introduced [251], [252], [253]. This method allows Alice to maintain confidentiality over the computation's input, process, and output. Assuming Alice can generate single qubit states, she can send these to Bob along with classical instructions for the computation without revealing the specific operations involved. Bob performs the computation as directed but without understanding the significance of his actions due to his ignorance of the initial state orientations set by Alice. Thus, Bob completes the computation without gaining any knowledge about the computation's specifics, effectively performing the computation "*blindly*". This approach safeguards the privacy of the computation, revealing only the necessary amount of information to Bob to perform the task without understanding its content.

## 5 CHALLENGES IN LONG DISTANCE COMMUNICATION

In the study of optical data transmission, considering the imperfections within a system becomes paramount. Achieving a hundred percent efficiency is difficult goal to achieve, and the output at the termination of an optical cable rarely mirrors the input. Consequently, a detailed examination of the primary contributors to loss in optical fibers is essential. Five distinct losses are broadly addressed: dispersion, absorption, scattering, bending, and coupling [254], [255], [256], [257].

### 5.1 Sources of Attenuation

*5.1.1. Mode Dispersion in Optical Fibers.* In multimode optical fibers, light propagation occurs through various modes, each following distinct paths with different speeds to cover the same distance '$L$' within the fiber. To comprehend the temporal disparities introduced by different modes, their respective paths are studied. Axial modes that travel along the central axis, take a direct route, while other modes follow longer, internally reflected paths. The axial mode, representing the fastest trajectory, and non-axial modes, characterized by total internal reflections, demand an understanding of refractive index. The refractive index ($n_f$) is a fundamental property in refraction influencing the speed of light in the fiber, dictating the temporal characteristics of mode propagation.

The minimum time delay, denoted as '$t_{min}$', corresponds to the axial mode's traversal of a certain length '$L$':

$$t_{min} = \frac{L}{v_f} \tag{88}$$

where $v_f = c/n_f$ is the speed of light in the fiber, $c$ is the speed of light in vacuum.

The maximum time delay, '$t_{max}$', occurs when considering a critical non-axial path sustaining total internal reflection:

$$t_{max} = \frac{l.\,n_f^2}{c.\,n_c} \tag{89}$$

where $l$ is the distance of the non-axial path, and $n_c$ is the refractive index of the cladding.

The temporal spread due to mode dispersion, quantified by $\Delta t$, is influenced by the distance traveled. The larger the distance, the greater the temporal spread:

$$\Delta t \propto t_{max} - t_{min} \tag{90}$$

This dispersion diminishes signal readability, imposing constraints on the frequency. To ensure readability the signal pulses should be more than twice the spread, which consequently constrains the bandwidth of the transmitted information.

*5.1.2. Absorption.* One pivotal contributor to signal attenuation is absorption, classified into intrinsic and extrinsic components. Intrinsic absorption, an inherent property of fibers, arises from the interaction between signal photons and electrons within the material. This interaction leads to electron excitation, resulting in diminished signal power. While intrinsic absorption is inevitable, its impact is comparatively minor. Conversely, extrinsic absorption emanates from impurities introduced during the fiber manufacturing process. This form of absorption is more substantial but controllable through meticulous manufacturing practices. Manufacturers typically endeavor to maintain impurity levels below 1% to mitigate extrinsic absorption's impact on signal attenuation.

*5.1.3. Scattering.* Scattering mirrors absorption in its dependence on impurities within the fiber. It involves the re-radiation and random re-emission of light in various directions. The manufacturing process plays a pivotal role in managing scattering by controlling impurity levels. Scattering encompasses diverse types, such as linear and non-linear scattering, which is beyond the current discussion's scope.

*5.1.4. Bending and Coupling.* Bending introduces another dimension to signal attenuation, especially when the physical curvature of the fiber disrupts the conditions for total internal reflection. Manufacturers stipulate minimum bending radii to circumvent signal loss. Coupling errors arise during the joining of two fibers that can result from gaps or misalignment, allowing signal leakage and contributing to attenuation.

### 5.2 Quantifying Attenuation

As signals traverse the fiber, they inevitably lose power. To quantify this power loss, the unit of decibels ($dB$) becomes instrumental [258], [259]. The decibel, representing the ratio of power out to power in, is expressed as a logarithmic function, as shown in the following expression.

$$dB = -10\log_{10}\frac{P_o}{P_i} \tag{91}$$

This logarithmic scale accommodates the wide-ranging power differentials encountered in optical fiber transmissions. A modification of equation 91 with an attenuation parameter, denoted as $\alpha$, signifies the number of decibels per kilometer.

$$\alpha = -\frac{10}{L}\log_{10}\frac{P_o}{P_i} \tag{92}$$

It is a critical metric in understanding how signals attenuate over distance. By manipulating the equation 92, one can gather insights into the relationship between power ratios, fiber length, and attenuation parameters, as shown below:

$$\frac{P_o}{P_i} = 10^{-\frac{\alpha L}{10}} \tag{93}$$

Real-time implementations provide clarity on the concept of attenuation parameters. For instance, in 1970, the attenuation was 20 dB/Km, reflecting a transmission of 1% of input power over one kilometer. Contrastingly, two decades later, advancements yielded an attenuation level of 0.18 dB/Km, facilitating a transmission of around 96% over the same distance.

### 5.3 Overcoming Losses in Optical Fibers

To mitigate dispersion, one effective strategy involves transitioning to a single-mode fiber, where only a single mode propagates. This shift eliminates the dispersion challenges associated with multimode fibers. Additionally, absorption and scattering losses, primarily stemming from impurities in the fiber, can be controlled through meticulous manufacturing processes. By minimizing impurities, both absorption and scattering effects are curtailed. Another critical aspect is minimizing bending-induced losses by avoiding unnecessary bends in the fiber. For coupling errors introduced during fiber junctions, proper alignment and the elimination of gaps between fibers play a significant role.

Despite these measures, some level of signal attenuation ($P_o$) is inevitable over a distance $L$ due to the attenuation parameter $\alpha$. This can be quantified by modifying the expression from equation 93:

$$P_o = P_i . 10^{-\frac{\alpha L}{10}} \tag{94}$$

where $P_i$ is the input power.

As the distance increases, even with low-loss fibers, the signal diminishes significantly. To address this, repeaters are crucial. Erbium-Doped Fiber Amplifiers (EDFAs) are particularly noteworthy in this regard [260], [261]. They leverage erbium atoms introduced into the fiber. Pumping these atoms creates population inversion, and the weak signal stimulates emission from excited erbium atoms, amplifying the signal. This amplification enables the signal to travel longer distances before requiring amplification again.

However, it's essential to be aware of repeater-induced noise. While EDFAs effectively amplify signals, they can also amplify noise. Spontaneous emissions from excited erbium atoms can lead to noise amplification, impacting the signal-to-noise ratio. Managing this ratio is critical for maintaining signal integrity in optical communication systems, especially those designed for long-distance signal transmission.

### 5.4 Quantum Challenges in Fiber Optics

In quantum communication, where signals exist at the granularity of individual photons, a distinctive challenge arises; the loss of these delicate particles within the fiber. Unlike classical communication, where losing a photon might be inconsequential, in quantum protocols like E91 or BB84, the loss of even a single photon results in protocol failure.

Addressing photon loss in quantum communication is not as straightforward as classical methods. While classical signals can be amplified, attempting the same for quantum signals encounters a significant hurdle known as the no-cloning theorem (see Section 4.3). This theorem prohibits the exact duplication of arbitrary quantum states, a fundamental barrier in creating backup copies for lost photons.

Considering alternative strategies, the idea of simply transmitting a photon and hoping for successful reception faces a probabilistic challenge. The transmission probability ($P_{transmit}$) through a fiber of length $L$, with attenuation parameter $\alpha$, can be computed from equations 93 and 94. Even in a best-case scenario with ultra-low attenuation (e.g., 0.1 dB/Km) over a long fiber (e.g., 1000 Km), the probability of successfully transmitting a single photon is exceedingly small i.e., $10^{-10}$. To emphasize, if a source emits one photon per second, it would take an average of 317 years for a distant receiver to successfully receive a single photon.

Beyond photon loss, long-distance quantum communication introduces additional challenges, including unitary errors like Pauli errors, along with non-unitary errors such as coherence, dephasing, and relaxation [262]. These quantum-specific challenges complicate the landscape, making it distinct from classical communication. The seemingly daunting scenario prompts exploration into quantum solutions for long-distance quantum communication challenges, a topic that will be further delved into in the next section.

## 6 QUANTUM REPEATERS: SCALING QUANTUM NETWORKS

In the quest for efficient long-distance quantum communication, the phenomenon of photon loss in fibers emerges as a significant challenge. The classical approach to address this involves direct connections between devices, forming a complete graph where each device connects to every other. However, this classical strategy faces scalability issues due to the need for specialized hardware for each connection.

Real-world networks adopt a more practical approach, avoiding the impracticality of all-to-all connections. Instead, they establish connections strategically, allowing communication between arbitrary nodes without an exhaustive network of dedicated links. Quantum communication introduces an intriguing approach to this challenge through the concept of quantum teleportation (refer Section 4.1 and 4.2).

Teleportation leverages entangled pairs between neighboring nodes, enabling the transfer of quantum information from the sender node to the target node by hopping through entangled intermediaries. However, this approach comes with a drawback in which the noisy nature of the operations and the entangled states degrades the quantum information significantly by the time it reaches the target node.

The ideal scenario involves having direct entangled pairs between communicating nodes, eliminating the need for extensive dedicated hardware and physical connections. Central to addressing these challenges is the concept of "entanglement". This involves establishing entanglement between neighboring nodes, forming a foundational element. The section aims to explore seven critical requirements: Firstly, it delves into the intricacies of "Link-Level Entanglement", investigating physical connections between neighboring nodes to establish the entanglement needed for quantum communication [263]. Secondly, it addresses the significant challenge of "Entanglement Swapping" [264]. This involves grappling with the complexities of extending entanglement over larger distances and multiple network hops without the luxury of dedicated physical connections. Thirdly, the section tackles the crucial aspect of "Detecting and Handling Sources of Error". This involves developing strategies to manage the inevitable noise [265], [266] and errors inherent in quantum processes. The focus then expands to "Network Management", exploring broader aspects such as routing, multiplexing, and resource management within quantum networks. Additionally, the discourse extends into "Multipartite Entanglements and Graphs" to understand the role of complex entanglement structures in network connectivity and optimization. The conversation also covers the "Practical Implementation of Bell-State Analyzer" as a cornerstone technology for detecting entanglement. Finally, "Quantum Error Correction" is explored in detail as an essential mechanism for preserving information integrity in the inherently noisy quantum environment.

## 6.1 Link-Level Entanglement

Initiating link-level entanglement involves two neighboring nodes equipped with quantum memory represented as grey cylinders in Figure 8. A Bell State Analyzer (BSA) is positioned at the network's core, forming a Memory-Interference-Memory (MIM) setup [267].

Let's consider an example where quantum memories of both nodes emit 2 sets of entangled photons, which undergo attenuation during fiber travel. Upon potential loss of the first photons, surviving photons reach the BSA for a Bell state measurement, as shown in Figure 8. The measurement, influenced by detector clicks and photon indistinguishability, projects quantum memories onto Bell pairs i.e., $|\Phi^+\rangle$, $|\Phi^-\rangle$, $|\Psi^+\rangle$ and $|\Psi^-\rangle$ (refer Section 2.2). This process establishes link-level entanglement between left and right node quantum memories. Detailed mathematical aspects will be explored in subsequent steps for distant network node entanglement.

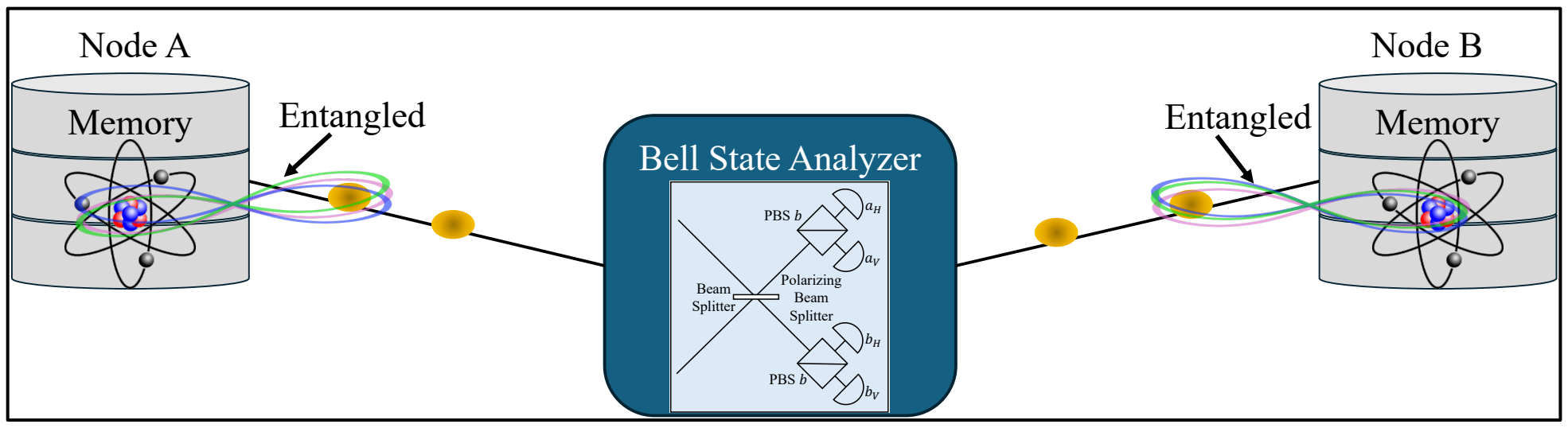

Figure 8. Link Level entanglement (MIM)

An alternative method involves direct memory-to-memory entanglement, viable in scenarios with minimal memory distance or exceptionally low attenuation fibers [268]. This approach necessitates a Bell state analyzer integrated into the network node for the Bell state measurement as shown in Figure 9.

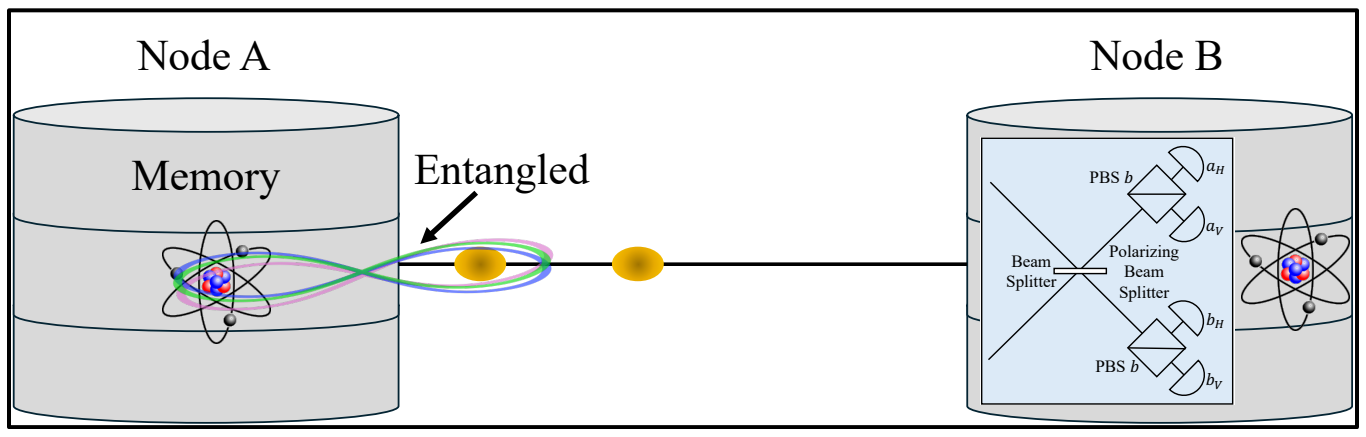

Figure 9. Link Level entanglement (MM)

## 6.2 Entanglement Swapping

The preceding subsection explored the establishment of link-level entanglement between neighboring nodes. This subsection delves into extending this entanglement over greater distances, achieving long-distance end-to-end entanglement [269]. Imagine three network nodes: 0, 1, and 2. Entanglement has been established between neighboring stations through specific qubit pairs. For clarity, let's designate the qubit at node 0 as $A$, and node 2 as $B$. Node 1 has 2 qubits i.e., $A'$ (entangled with $A$ at node 0) and $B'$ (entangled with $B$ at node 2). The objective is to establish maximal entanglement between nodes 0 and 2 via qubits $A$ and $B$.

Let's assume that the nodes 0 and 1 are entangled via a Bell pair $|\Psi^+\rangle$ and nodes 1 and 2 are entangled via the Bell pair $|\Phi^-\rangle$. Both entanglements are shown in the mathematical expressions given below:

$$|\Psi^+\rangle_{01} = \frac{1}{\sqrt{2}}(|01\rangle + |10\rangle);\ |\Phi^-\rangle_{12} = \frac{1}{\sqrt{2}}(|00\rangle - |11\rangle) \tag{95}$$

The combined state due to the tensor product of $|\Psi^+\rangle_{01}$ and $|\Phi^-\rangle_{12}$ is:

$$|Combined\ State\rangle = |\psi\rangle = |\Psi^+\rangle_{01} \otimes |\Phi^-\rangle_{12} = \frac{1}{2}(|0100\rangle - |0111\rangle + |1000\rangle - |1011\rangle) \quad (96)$$

Manipulating Equation 96 to convert it to 2-qubit basis using equations 57-60, the following can be derived:

$$|\Psi^+\rangle_{01} \otimes |\Phi^-\rangle_{12} = \frac{1}{2}(|0100\rangle - |0111\rangle + |1000\rangle - |1011\rangle) = \frac{1}{2}(|0\rangle|10\rangle|0\rangle - |0\rangle|11\rangle|1\rangle + |1\rangle|00\rangle|0\rangle - |1\rangle|01\rangle|1\rangle)$$
$$= \frac{1}{2\sqrt{2}}(|0\rangle\,(|\Psi^+\rangle - |\Psi^-\rangle)\,|0\rangle - |0\rangle\,(|\Phi^+\rangle - |\Phi^-\rangle)\,|1\rangle\; + |1\rangle\,(|\Phi^+\rangle + |\Phi^-\rangle)\,|0\rangle - |1\rangle\,(|\Psi^+\rangle + |\Psi^-\rangle)\,|1\rangle)$$
$$= \frac{1}{2\sqrt{2}}((|\Psi^+\rangle - |\Psi^-\rangle)|0\rangle|0\rangle + (|\Phi^+\rangle - |\Phi^-\rangle)|0\rangle|1\rangle + (|\Phi^+\rangle + |\Phi^-\rangle)|1\rangle|0\rangle - (|\Psi^+\rangle + |\Psi^-\rangle)|1\rangle|1\rangle)$$
$$= \frac{1}{2\sqrt{2}}((|\Psi^+\rangle - |\Psi^-\rangle)|00\rangle + (|\Phi^+\rangle - |\Phi^-\rangle)|01\rangle + (|\Phi^+\rangle + |\Phi^-\rangle)|10\rangle - (|\Psi^+\rangle + |\Psi^-\rangle)|11\rangle)$$
$$= \frac{1}{2\sqrt{2}}(|\Psi^+\rangle\,(|00\rangle - |11\rangle) - |\Psi^-\rangle\,(|00\rangle + |11\rangle) + \;|\Phi^+\rangle(|10\rangle - |01\rangle) + \;|\Phi^-\rangle(|10\rangle + |01\rangle)) \quad (97)$$

In Equation 97, the qubits at Node 1 are measured. Therefore, the manipulation of the combined state is focused around representing the second and third qubits as combinations of indistinguishable Bell States. In the previous derivation, it has been shown that the if the Bell State Measurement is conducted at Node 1, then the qubits at Nodes 0 and 2 are projected onto a maximally entangled Bell State. For example, equation 97 shows that if the measurement at Node 1 produces a result $|\Psi^+\rangle$ then the qubits at Nodes 0 and 2 are projected on to state $|\Phi^-\rangle$ and so on.

However, it's essential for Node 1 to communicate the measurement outcomes to both Node 0 and Node 2. This classical notification introduces a time delay in advancing entanglement establishment between the distant nodes.

Comparatively, the entanglement swapping procedure parallels the link-level entanglement. Mathematically, there's similarity in the swapping of entanglement between photons and memories. Physically, the distinction lies in the use of linear optics, limiting the success probability in photon entanglement swapping to 50%. Conversely, deterministic entanglement swapping between memories is achievable with precise experimental techniques and mitigating noise effects. An alternative perspective involves understanding entanglement swapping through teleportation (refer Section 4.2) [264], [269]. This conceptualization involves sharing a maximally entangled state between qubits $B'$ and $B$, with a Bell state measurement effectively teleporting the state from $A'$ to $B$. This method illuminates the process of transferring entanglement between qubits across the network.

### 6.3 Detecting and Handling Sources of Error

The primary objective is to create a state between station zero and station two, ensuring its quality is acceptable. Let the aimed state is the maximally entangled state $|\Phi^+\rangle$. However, real-world operations are inevitably accompanied by errors, introducing noise into the goals. The actual state, denoted by $\rho$ is a mix of the desired state $\Phi^+$ and noise terms, expressed as:

$$\rho = F|\Phi^+\rangle\langle\Phi^+| + (1 - F)|Noise\rangle\langle Noise| \quad (98)$$

Here, $F$ is the fidelity, representing the probability of obtaining the pure state $\Phi^+$, and$(1 - F)$ accounts for the probability of encountering noise. The challenge is to discern and rectify these imperfections.

One example of an error-inducing channel is the bit flip channel, symbolized by the Pauli $X$ matrix. In the context of the scenario, it affects the Bell state $\Phi^+$. The bit flip channel transforms the state $\Phi^+$ into $\Psi^+$, where $\Psi^+$ is the result of applying the $X$ matrix to one of the qubits of $\Phi^+$. Equation 98 can thus be modified into the following:

$$\rho = F|\Phi^+\rangle\langle\Phi^+| + (1 - F)|\Psi^+\rangle\langle\Psi^+| \quad (99)$$

To tackle errors, a simplified approach is explored called "*purification*" [270]. This method involves the exchange of classical information between distant nodes, Node 0 and Node 2, without compromising the entanglement.

The purification process employs another set of entangled pairs, referred to as Bell pairs [271]. Both Alice and Bob, situated in Node 0 and Node 2, respectively, manipulate their qubits through local operations. The timeline of these operations is synchronized to ensure optimal purification. Let both the entangled qubits are represented as follows:

$$Qubit\ 1: |\Phi^+\rangle_1 = \frac{1}{\sqrt{2}}(|00\rangle + |11\rangle);\ Qubit\ 2: |\Phi^+\rangle_2 = \frac{1}{\sqrt{2}}(|00\rangle + |11\rangle)$$

Then the combined state can be expressed as follows:

$$|\Phi^+\rangle_1|\Phi^+\rangle_2 = \frac{1}{2}(|00\rangle + |11\rangle)(|00\rangle + |11\rangle) = \frac{1}{2}[|00\rangle|00\rangle + |00\rangle|11\rangle + |11\rangle|00\rangle + |11\rangle|11\rangle]$$

$$CNOT_{Bob}\left[CNOT_{Alice}[|\Phi^+\rangle_1|\Phi^+\rangle_2]\right] = CNOT_{Bob}\left[\frac{1}{2}[|00\rangle|00\rangle + |00\rangle|11\rangle + |11\rangle|10\rangle + |11\rangle|01\rangle]\right]$$

$$= \frac{1}{2}[|00\rangle|00\rangle + |00\rangle|11\rangle + |11\rangle|11\rangle + |11\rangle|00\rangle] = |\Phi^+\rangle_1|\Phi^+\rangle_2$$

It can be seen that the same state has been retained after the application of CNOT gates on Alice's and Bob's side. Therefore, after applying CNOT gates and subsequent measurements, Alice and Bob share their results. If the measurements are correlated, indicating agreement, they retain the first pair of qubits. This process ensures the preservation of the entangled state while obtaining information about the fidelity of the initial Bell pairs [272]. In the ideal scenario, where no errors occur, the success probability of this purification process is:

$$\mathbb{P}_{success} = F^2 + (1 - F)^2 \tag{100}$$

This leads to the final fidelity of:

$$F' = \frac{F^2}{\mathbb{P}_{success}} \Rightarrow F' = \frac{F^2}{F^2 + (1 - F)^2} \tag{101}$$

Purification offers a valuable tool for enhancing the fidelity of entangled states [270]. Especially, when the initial fidelity surpasses $F > 0.5$, the purification process acts to increase the fidelity of the resulting pairs. This iterative application of purification can progressively elevate the quality of end-state fidelities, exemplifying its apt designation.

### 6.4 Network Management

In the pursuit of building a quantum network, the discussion till now has progressed from establishing entanglement between neighboring nodes, extending this to end-to-end entanglement through entanglement swapping. This leads to the requirement of delving into the challenges of creating a functional network, addressing routing, multiplexing, and resource management.

Routing in a quantum network involves determining the path for establishing entangled pairs between nodes [273]. This requires evaluating the costs associated with link-level entanglement. The cost is measured in "*Seconds per Bell pair at a specified Fidelity Threshold*" [274]. The reason for such a distinction is rooted in the time taken to establish a link-level entanglement. The photon attenuation/loss, in general, is probabilistic in nature. Therefore, a successful establishment of entanglement is probabilistic as well. Factors such as channel loss impact the time taken to establish entanglement, influencing the overall cost. One of the popular routing methods is Dijkstra's shortest path algorithm [275], [276], which aids in efficiently combining these costs for various links, facilitating the determination of the most cost-effective route for end-to-end entanglement.

To exemplify the concept of routing in a quantum network, let's consider a simple illustrative example with five nodes, as shown in Figure 10. Given the quantum network setup with nodes $A$, $B$, $C$, $D$, and $E$, and the direct link costs in *seconds per Bell pair at a specified fidelity threshold*, the weighted adjacency matrix $M$ for this scenario can be constructed as follows. Let's define the matrix such that each cell $(i, j)$ in the matrix represents the cost of establishing entanglement between nodes $i$ and $j$. If two nodes are not directly connected, their corresponding cell will have a value of ∞ to indicate the absence of a direct link.

$$M = \begin{bmatrix} 0 & 4 & 6 & \infty & \infty \\ 4 & 0 & 3 & 5 & \infty \\ 6 & 3 & 0 & 2 & \infty \\ \infty & 5 & 2 & 0 & 1 \\ \infty & \infty & \infty & 1 & 0 \end{bmatrix}$$

This matrix $M$ represents the time it takes to establish a high-fidelity entangled pair between directly connected nodes.

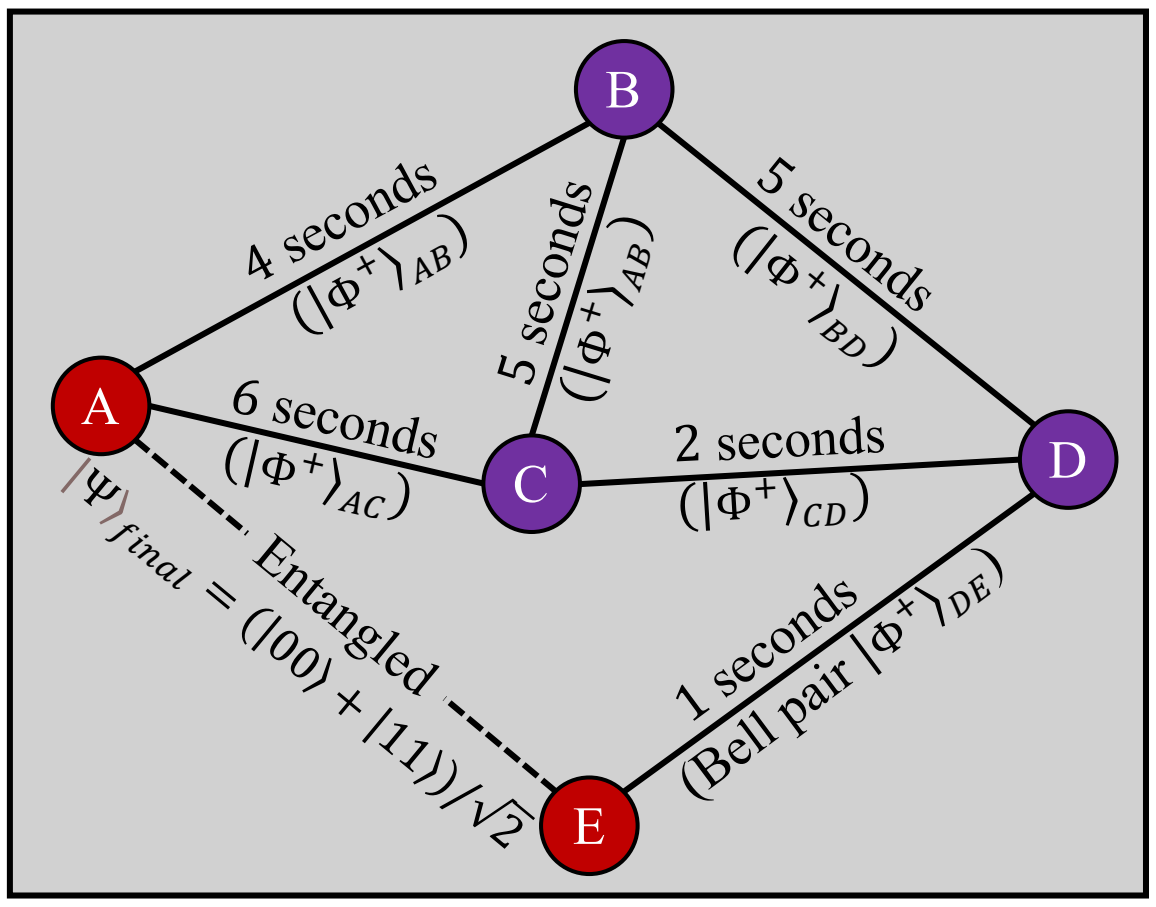

Figure 10. Routing in Quantum Networks

The objective is to identify the most efficient path for establishing entanglement between $A$ and $E$, considering the time to establish a high-fidelity entangled pair across multiple nodes. To demonstrate the Dijkstra distance update steps for quantum routing, the cost computation starts from node $A$ and aim to find the shortest path to all other nodes in the network ($B$, $C$, $D$, and $E$), based on the weighted adjacency matrix provided. The goal is to establish entanglement with $E$, considering the minimum total time, which is demonstrated in Table 3. This process effectively illustrates how Dijkstra's algorithm functions in determining the shortest path for quantum entanglement distribution across a network, taking into account the "*Seconds per Bell pair at a specified Fidelity Threshold*" as the link cost metric.

Table 3: Example Implementation of Dijkstra's Algorithm

| Step | Current Node | Distance to A | Distance to B | Distance to C | Distance to D | Distance to E | Path |
|---|---|---|---|---|---|---|---|
| **0** | Initialization | 0 | ∞ | ∞ | ∞ | ∞ | A |
| **1** | A | 0 | 4 | 6 | ∞ | ∞ | A |
| **2** | B | 0 | 4 | 7 | 9 | ∞ | A→B |
| **3** | C | 0 | 4 | 7 | 5 | ∞ | A→B→C |
| **4** | D | 0 | 4 | 7 | 5 | 6 | A→B→C→D |

In practice, establishing entanglement across this path involves initial entanglement between directly connected nodes ($AB$, $BC$, $CD$ and $DE$). This is followed by sequential entanglement swapping at intermediate nodes ($B$, $C$ and $D$) to extend entanglement from $A$ to $E$ (refer Section 6.3). The process of entanglement swapping involves performing operations on the entangled pairs to extend entanglement across the network. Let's say that the entangled pair shared by adjacent nodes in the network is at entangled state $|\Phi^+\rangle$. The quantum state of the entire network before any swapping can be represented as a tensor product of the entangled pairs between adjacent nodes:

$$|\Psi\rangle_{initial} = |\Phi^+\rangle_{AB} \otimes |\Phi^+\rangle_{BC} \otimes |\Phi^+\rangle_{CD} \otimes |\Phi^+\rangle_{DE}$$

Given that a Bell State Measurement (BSM) is performed at nodes $B$, $C$ and $D$ for swapping, after successive BSMs and entanglement swapping, $A$ and $E$ share an entangled state. The final state $|\Psi\rangle_{final}$ between $A$ and $E$ mirrors the initial entangled pairs, now spanning the network:

$$|\Psi\rangle_{final} = \frac{1}{\sqrt{2}}(|00\rangle + |11\rangle) = |\Phi^+\rangle$$

This scenario demonstrates the foundational quantum mechanics and Hilbert space mathematics underlying quantum routing and entanglement distribution in a quantum network.

Although being able to accomplish routing via end-to-end entanglement establishment, multiplexing in quantum networks presents a challenge. This is experienced when multiple nodes seek to establish end-to-end entanglement, creating contention on shared links. For instance, if there are some common nodes and links in a network which are part of multiple end-to-end entanglements. Efficient resource management becomes crucial for handling these scenarios, optimizing link usage and resolving contention. These considerations are crucial for ensuring efficient and reliable end-to-end entanglement establishment within the network. Building a quantum network is a multifaceted endeavor that extends beyond the basics of entanglement creation, involving intricate strategies for network management and optimization.

### 6.5 Multipartite Entanglements and Graphs

Establishing a link-level entanglement or an end-to-end entanglement between two nodes is necessary for a conceptual understanding of quantum communication. However, tacking standard requisites in a communication network like multiplexing, routing, contention-free link usage etc. poses as challenges in ensuring the creation of a realizable network. Therefore, handling communication in a quantum network requires going beyond the concepts of bipartite entanglement. Let's take the case of a tripartite entanglement, which is entanglement shared between 3 nodes. Such an entanglement will have 8 basis states i.e., $|000\rangle$, $|001\rangle$, $|010\rangle$, $|011\rangle$, $|100\rangle$, $|101\rangle$, $|110\rangle$ and $|111\rangle$. So, following the idea of superposition, any 3-qubit general state can be written as a superposition of all of these basis states. This is shown in the following expression:

$$|\psi\rangle = c_1|000\rangle + c_2|001\rangle + c_3|010\rangle + c_4|011\rangle + c_5|100\rangle + c_6|101\rangle + c_7|110\rangle + c_8|111\rangle \qquad (102)$$

Here, $c_1$ to $c_8$ are the probability amplitudes of the basis states. Now, based on the characteristics of the superposition the state $|\psi\rangle$ can be a separable state or an entangled state. If a particular state is considered with equal superposition of basis states $|000\rangle$ and $|111\rangle$, it can be shown as:

$$|GHZ\rangle = \frac{1}{\sqrt{2}}(|000\rangle + |111\rangle) \qquad (103a)$$

This state is called a *Greenberger–Horne–Zeilinger* state ($GHZ$-state) which forms the basis for *M-partite system*, where $M$ is the number of qubits used in the system. In Eqn. 103, $M = 3$ as it represents a 3-qubit system. If a measurement is conducted for one of the qubits of $GHZ$-state in $Z$-basis, the outcomes are as follows:

$$Prob\{+1\} = Prob\{-1\} = \frac{1}{2} \qquad (103b)$$

The probabilities from Eqn. 103b shows that with an outcome of $+1$, the state of one qubit is projected to state $|0\rangle$, consequently collapsing the state of three qubits to state $|000\rangle$. Similarly, an outcome of $-1$ collapses the 3-qubit state to $|111\rangle$. This manifests similar behavior as described in "Entanglement-based QKD", shown in Eqns. 80-83, where the

outcomes of the qubit measurements are correlated to each other. The physical meaning of this is that there exists an entanglement between three parties. To be noted that, similar to the measurement of bipartite measurement, the entanglement is destroyed post-measurement.

Let's consider another 3-qubit state given by the following:

$$|W\rangle = \frac{1}{\sqrt{3}}(|001\rangle + |010\rangle + |100\rangle) \qquad (104a)$$

Here, the measurement of the first qubit in $Z$-basis will result in unequal probabilities, which is shown below:

$$Prob\{+1\} = \frac{2}{3}, \qquad Prob\{-1\} = \frac{1}{2} \qquad (104b)$$

When the first qubit is projected on the state $|0\rangle$, the remaining two qubits remain in a maximally entangled state $|\Psi^+\rangle = \frac{1}{\sqrt{2}}(|01\rangle + |10\rangle)$ (refer Eqn. 55). Unlike $GHZ$-state, the initial entanglement in $W$-state is not destroyed entirely. This brings about the discussion on the level of correlation among qubits in an entangled state. According to the concept of "*Monogamy of Entanglement*", is two qubits are strongly correlated then their level of correlation with a third qubit is limited.

This fundamental can be extended to *M-partite* $GHZ$-state and $W$-state. The mathematical expressions are:

$$|GHZ\rangle = \frac{1}{\sqrt{2}}(|00 \ldots 0\rangle + |11 \ldots 1\rangle) \qquad (105a)$$

$$|W\rangle = \frac{1}{\sqrt{M}}(|00 \ldots 1\rangle + |01 \ldots 0\rangle + \cdots + |10 \ldots 0\rangle) \qquad (105b)$$

Note that Eqn. $105a$ has two terms and Eqn. $105b$ has $M$ terms.

Another scenario that manifests multipartite entanglement, is called the *graph state* $|G\rangle$ [277], [278], [279]. Here, a control-$Z$ gate is used to establish entanglement between all qubits by controlled application of $Z$ gate. To be noted that the control-$Z$ gate is similar to $CNOT$ gate, where Pauli-$Z$ gate is applied if the control bit is in state $|1\rangle$. A graph state with regular lattice structure is also called *cluster state*, $|C\rangle$ [280], [281], [282]. *M-partite* states have found their place in measurement-based quantum computation, quantum error correction etc. One of the most accepted uses of *M-partite* states is in Secret sharing, which is a procedure for splitting a message into several parts so that no subset of parts is sufficient to read the message.

### 6.6 Practical Implementation of Bell-State Analyzer

Till now this article has covered the fundamentals of entanglement, its utility in secured communication and how they are leveraged in quantum repeaters. Given the uncertainty associated with determining the entangled states, one of the crucial steps encountered is the "*Measurement of Bell States*". The theoretical foundation of bell states measurement is grounded on the application "*Born Rule*" to determine "*which of the four Bell states is the given state in?*".

To show the implementation of Bell states measurements, let's take a generalized expression where any state in 2-qubit computational basis is converted into Bell basis. This is shown in the below expression where Eqn. 8 is represented in terms of bell states $|\Phi^+\rangle$, $|\Phi^-\rangle$, $|\Psi^+\rangle$ and $|\Psi^-\rangle$ using equations 53-60:

$$|\psi\rangle = \alpha|00\rangle + \beta|01\rangle + \gamma|10\rangle + \delta|11\rangle = \left(\frac{\alpha+\beta}{\sqrt{2}}\right)|\Phi^+\rangle + \left(\frac{\alpha-\beta}{\sqrt{2}}\right)|\Phi^-\rangle + \left(\frac{\gamma+\delta}{\sqrt{2}}\right)|\Psi^+\rangle + \left(\frac{\gamma-\delta}{\sqrt{2}}\right)|\Psi^-\rangle \quad (106)$$

Now, following the Born rule (refer Eqn. 62) the measurement can be conducted in Bell basis, i.e., $|\langle Bell\ state|\psi\rangle|^2$, which will result in one of the Bell states. This is a very abstract notion of bell state measurement.

A practical implementation of bell state measurement in *Bell State Analyzer* is different than the above-stated abstraction [283]. It requires the measurement to be conducted in 2-qubit $Z$-basis. Before understanding the procedure to used two Pauli-Z measurements to implement Bell state measurement, let's see a 1-qubit Pauli-X measurement in terms of Pauli-Z measurement. Pauli-X measurement of a state $|\psi\rangle = \alpha|0\rangle + \beta|1\rangle$ can be conducted in two ways. The first method is directly conducting Pauli-X measurement using Eqn. 2:

$$|\psi\rangle = \alpha|0\rangle + \beta|1\rangle = \frac{\alpha+\beta}{\sqrt{2}}|+\rangle + \frac{\alpha-\beta}{\sqrt{2}}|-\rangle \quad (107a)$$

$$Prob\{+1\} = |\langle\psi| +\rangle|^2 = \frac{1}{2}(\alpha+\beta)^2 \quad (107b)$$

$$Prob\{-1\} = |\langle\psi| -\rangle|^2 = \frac{1}{2}(\alpha-\beta)^2 \quad (107c)$$

An alternate method is to use Hadamard gate (see Eqn. 21) to the state $|\psi\rangle$ and then conducting the measurement in $Z$ basis. Applying Hadamard gate to $|\psi\rangle$ converts the $Z$ bases to $X$ bases i.e., $|+\rangle$ and $|-\rangle$. The converted state $|\psi'\rangle$ in $X$ basis is shown below:

$$|\psi'\rangle = H|\psi\rangle = H(\alpha|0\rangle + \beta|1\rangle) = \alpha|+\rangle + \beta|-\rangle \quad (108a)$$

Now to conducting a Pauli-Z measurement:

$$|\psi'\rangle = \alpha|+\rangle + \beta|-\rangle = \frac{\alpha+\beta}{\sqrt{2}}|0\rangle + \frac{\alpha-\beta}{\sqrt{2}}|1\rangle \quad (108b)$$

$$Prob\{+1\} = |\langle\psi'|0\rangle|^2 = \frac{1}{2}(\alpha+\beta)^2 \quad (108c)$$

$$Prob\{-1\} = |\langle\psi'|1\rangle|^2 = \frac{1}{2}(\alpha-\beta)^2 \quad (108d)$$

This shows that Hadamard gate is the unitary operation that is required to conduct $X$-basis measurement using $Z$-basis [284], [285].

Now extending this concept to 2-qubit $Z$-basis measurement of Bell states, a 2-qubit unitary operation is required which can convert the Bell basis to the computation basis i.e., $|00\rangle$, $|01\rangle$, $|10\rangle$ and $|11\rangle$. This operation can be achieved by a Controlled NOT ($CNOT$) gate. This entails Hadamard gate operation on the first qubit followed by two measurements in $Z$-basis (see Figure 11). Each outcome $|00\rangle$, $|01\rangle$, $|10\rangle$ and $|11\rangle$ of these measurements will correspond to Bell states $|\Phi^+\rangle$, $|\Psi^+\rangle$, $|\Phi^-\rangle$ and $|\Psi^-\rangle$, respectively.

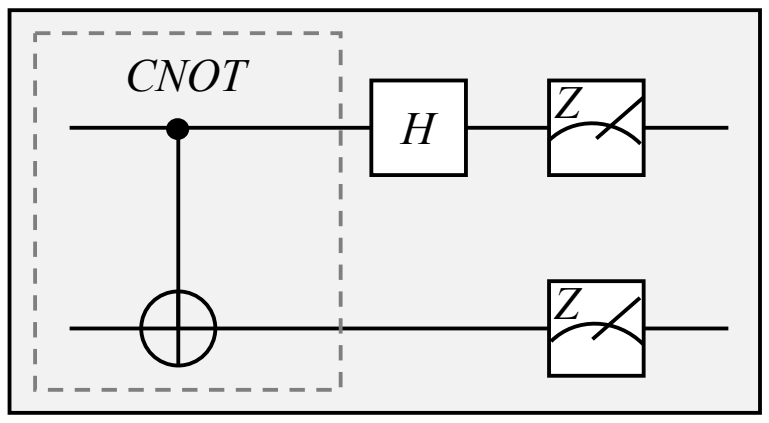

Figure 11. BSM using Pauli $Z$ measurement.

To be noted that Bell state measurement is not always successful in linear optics since it has been experimentally proven that the success probability is limited to 50% [286], [287]. This adds to the contention in the implementation of quantum repeaters and their limitations in the establishing entanglement between two nodes in a quantum network. It was discussed in the previous subsections that the quantum repeaters are subjected to noise while establishing entanglement between nodes and state purification is conducted to counteract the effect of noise. Additionally, there are the instances of attenuation due to lossy fibers. It must be emphasized that even after the successful mitigation of detrimental effects due to lossy fibers and noise, there lies a fundamental limitation in Bell state measurement.

### 6.7 Quantum Error Correction

Error correction in quantum networks is critical due to the inherently fragile nature of quantum states and the challenges posed by decoherence, quantum noise, and operational errors. These factors can lead to the corruption of quantum information during storage, processing, and transmission, undermining the reliability and functionality of quantum networks. A brief on the key effects of such challenges and the requirement for error correction in quantum networks is stated here.

Quantum systems operate on the principles of superposition and entanglement, yet these critical quantum properties are at risk of degradation through a phenomenon known as decoherence. *Decoherence* occurs when a quantum system, unavoidably interacting with its surrounding environment, begins to lose its quintessential quantum characteristics. Moreover, qubits are vulnerable to various disturbances, often referred to as quantum noise. These may stem from a range of sources including, but not limited to, imperfect control signals, thermal fluctuations, or general environmental interference.

These errors are not merely confined to the static storage of quantum information; they extend to dynamic processes as well. When quantum operations or gates are executed, any lack of precision in control mechanisms or any slight imperfection in the quantum hardware can lead to errors. Over time, as these errors accumulate, they can lead to substantial deviations from the desired outcomes of quantum computations.

Within the broader scope of quantum networks, the above-stated challenge intensifies. Quantum information often needs to be transmitted across vast distances, a task typically achieved through optical fibers or even free space. However, such transmission is not immune to losses, scattering, and various forms of degradation that can compromise the integrity of the quantum states being conveyed. Considering these challenges, the scalability of quantum networks and their ability to support intricate and advanced quantum computing and communication tasks hinge on their fidelity and robustness. Maintaining a high degree of accuracy in quantum state manipulation and preventing errors are thus paramount to the success and reliability of quantum technologies.

Quantum error correction (QEC) is a critical facet of quantum information science, aimed at preserving the coherence of quantum states against the detrimental effects of environmental noise and operational inaccuracies. QEC schemes are designed to enable the detection and rectification of quantum errors while circumventing the collapse of the quantum state due to direct measurement [288]. Following is a technical overview of a subset out of various existing QEC codes and the underlying mathematical frameworks.

*6.7.1. Shor Code.* The Shor code, an important QEC protocol, is capable of rectifying arbitrary single-qubit errors such as bit flips ($X$), phase flips ($Z$), or their combination ($Y$) [289]. This is achieved by encoding a single logical qubit into a block of nine physical qubits, utilizing a concatenated code structure that combines three-qubit bit-flip correction with three sets of such triplets to handle phase-flip errors [253], [254], [255], [256]. A logical qubit can be defined as a protected form of quantum information that is encoded across multiple physical qubits using quantum error correction codes [254], [255], [256]. The following is a mathematical expression showing a logical qubit and its Shor code encoding:

$$|\psi\rangle = \alpha|0\rangle + \beta|1\rangle \Rightarrow |\psi\rangle_{encoded} = (\alpha|000\rangle + \beta|111\rangle)^{\otimes 3} \tag{109}$$

The purpose of a logical qubit is to reduce the effective error rate by distributing the information in such a way that local errors can be detected and corrected without disturbing the quantum information.

*6.7.2. Steane Code.* The [[7,1,3]] CSS (Calderbank-Shor-Steane) code, known as the Steane code, encodes one logical qubit into seven physical qubits [250], [257], [258], [259]. It arises from the amalgamation of classical error-correcting codes, specifically the [7,4,3] Hamming code, and features the property of being able to correct single-qubit errors by performing a special type of measurement called syndrome measurements which does not disturb the quantum information and that discern bit and phase flip errors separately, thereby preserving the encoded quantum information [299], [300]. The logical qubit states $|0\rangle_L$ and $|1\rangle_L$ are represented by the superposition of all even weight codewords and their orthogonal complements, respectively.

$$|0\rangle_L = \frac{1}{\sqrt{8}}\begin{pmatrix} |0000000\rangle + |1010101\rangle + |0110011\rangle + \\ |1100110\rangle + |0001111\rangle + |1011010\rangle + \\ |0111100\rangle + |1101001\rangle \end{pmatrix};$$

$$|1\rangle_L = X_L|0\rangle_L \tag{110}$$

Here, $X_L = X \otimes X \otimes\ X \otimes\ X \otimes X \otimes X \otimes X$, that is $X$ gate for each of 7 qubits.

*6.7.3. Surface Codes.* Surface codes represent a robust class of topological QEC codes that facilitate qubit encoding into a two-dimensional lattice structure [289], [301], [302], [303]. These codes are characterized by a high threshold for errors and enable the local measurement of error syndromes using stabilizer operators. Stabilizers here are products of Pauli $X$ and $Z$ operators acting on qubits that form vertices (to detect phase errors) and faces (to detect bit errors) of the lattice. The error syndromes are measured by the stabilizer operators, which for a simple surface code can be represented as:

$$Z_{stabilizers} = Z_i Z_j Z_k Z_l;\ X_{stabilizers} = X_i X_j X_k X_l \tag{111}$$

where $Z$ and $X$ refer to *Pauli* operators applied to the qubits $i$, $j$, $k$ and $l$ that form the corners of a local region in the lattice. The locality of these stabilizer checks is conducive to practical implementation in quantum processors [304].

*6.7.4. Quantum Reed-Muller Codes.* Quantum Reed-Muller codes are adapted from their classical counterparts and are known for their ability to correct both bit and phase errors [305], [306], [307], [308]. The encoding of a logical qubit into multiple physical qubits is executed via a transformation that disperses the qubit information, allowing for error correction by evaluating parity checks across various qubit subsets.

*6.7.5. Cat Codes.* Cat codes encode quantum information into coherent state [309] superpositions within a harmonic oscillator. These codes are tailored to counteract photon loss by representing logical qubits as superpositions of coherent states $\alpha$ and $-\alpha$, where $\alpha$ is a complex number representing the amplitude and phase of the coherent states. For example, a logical $|0\rangle_L$ could be encoded as $|\alpha\rangle + |-\alpha\rangle$ and $|0\rangle_L$ as $|\alpha\rangle - |-\alpha\rangle$. They leverage the non-orthogonality of coherent states to detect errors without destroying the quantum information.

*6.7.6. Bacon-Shor Codes.* These are a family of quantum error correction codes that simplify the correction process by requiring only two-qubit interactions. They are particularly aimed at correcting errors in a specific model of quantum computation where certain types of errors are more likely than others. These codes use a rectangular array of qubits and correct for arbitrary single-qubit errors through a simpler, more practical approach. In a simplified version, a $3 \times 3$ Bacon-Shor code encodes 1 logical qubit into 9 physical qubits, with the logical qubit states represented as certain symmetries within the array. The code stabilizers are products of $X$ or $Z$ operators along rows and columns to detect and correct errors. They are specifically oriented towards systems with asymmetric noise profiles and enable the correction of arbitrary single-qubit errors by leveraging symmetries and redundancy within the qubit array.

*6.7.7. General Mathematical Framework for QEC.* Quantum Error Correction codes work by entangling the logical qubit(s) with additional ancillary qubits, creating an enlarged Hilbert space where errors are detectable via stabilizer and syndrome measurement [310]. The general mathematical framework involves:

*a*) Logical qubits: Represented by $|\psi\rangle$, the state we wish to protect, like in Eqns. 109 and 110.

*b*) Physical qubits: The larger set of qubits that include the logical and ancillary qubit(s) for error detection.
*c*) Stabilizers: Operators that commute with the Hamiltonian of the system and are used to detect errors without measuring the logical qubits directly.
*d*) Syndrome measurement: The process of measuring the stabilizers to detect the occurrence and type of an error.

The objective of quantum error correction is to apply a correction operation that returns the system to its intended state without directly measuring or disturbing the state itself. This is represented mathematically by encoding and decoding operations, along with syndrome measurement and correction operations that depend on the specific code being used.

The QEC techniques highlighted above represent just a selection from a broader and rapidly evolving field. Researchers continue to build upon these foundational methods, developing advanced and novel QEC strategies that not only enhance fault tolerance for quantum computing but also significantly improve quantum communication. These advancements are paving the way toward more robust quantum systems, capable of maintaining coherence over longer periods and across greater distances, thus bringing the reality of practical and scalable quantum computing and communication ever closer.

## 7 QUANTUM NETWORKS OF NETWORKS

The concept of Quantum networks of networks, or more broadly the *Quantum Internet*, signifies a groundbreaking advancement in information technology, utilizing the unique properties of quantum mechanics to revolutionize communication, computing, and sensing across the globe. At the core of quantum networks lie previously discussed foundational principles such as quantum superposition, entanglement and teleportation, which enable novel forms of information transmission and processing. The research in [311] highlights the computational advantages of interconnecting quantum devices via a quantum internet. It delves into the foundational aspects of quantum computing, such as the role of qubits and the principle of superposition, and emphasizes the importance of entanglement and quantum teleportation for distributed quantum computing. The study also addresses the challenges associated with scaling up quantum computing power and the physical constraints on qubit connectivity, pointing out the critical need for efficient mapping between quantum algorithms and physical architectures. To address such challenges, a study titled FENDI [312] explores noise and fidelity issues in quantum networks, focusing on the probabilities of entanglement generation and swapping success. It models the effect of unsuccessful operations and inaccurate operations or noisy channels on the fidelity of entangled bits (ebits), highlighting the importance of optimizing these factors for efficient quantum communication. A *qudit* simulator called QuantumSkynet [313] focuses on cloud-based architecture for simulating high-dimensional quantum computing, highlighting the implementation of quantum phase estimation and Deutsch–Jozsa algorithms in high dimensions. The simulator demonstrates the potential for increased information encoding and resilience against decoherence.

The infrastructure of a quantum network includes several key technological components, one of which is Quantum repeaters that extend the range of quantum communication by performing entanglement swapping and purification, acting as the backbone for long-distance quantum connectivity. Quantum routers leverage the attributes of quantum repeaters to direct qubits efficiently across the network, while maintaining their quantum states. The work in [314] explores the concept of quantum repeaters as essential components for extending the range of quantum key distribution (QKD) and other cryptographic protocols, thereby securing global communication. It underscores the shift towards a Second Quantum Revolution, highlighting the experimental progress in quantum computing and the potential for secure access to cloud-based quantum computing through a quantum internet. The study also discusses the role of quantum networking in distributed quantum computing and the crucial use of photons as information carriers in quantum communication.

The applications of quantum networks are vast and varied, some of which have been discussed in the paper. Quantum Key Distribution (QKD) offers a secure method for distributing encryption keys, protected by the principles of quantum mechanics. Implementations based on the Software-Defined Network (SDN) paradigm, such as the Spanish network MadQCI, have been shown in [315]. It shows promise in managing both classical and quantum elements of quantum communication networks (QCNs). This approach enhances network flexibility, scalability, and supports the integration of QKD devices from different providers. Distributed quantum computing, facilitated by the interconnection of quantum computers through the network, promises computational power far beyond classical capabilities. The network also promises to revolutionize secure communication and enhance precision in timing and sensing technologies, offering accuracy unattainable by current systems.

Recent advances in quantum satellites have significantly propelled the capabilities for global quantum communication, overcoming terrestrial limitations. One noteworthy milestone is the launch of China's quantum satellite, Micius, which has executed long-distance entanglement distribution, demonstrating quantum key distribution (QKD) over distances that far surpass those achievable with optical fibers [316]. Micius achieved entanglement between two ground stations over 1200 km apart, and facilitated secure communication through QKD protocols like decoy-state BB84, enabling an ultra-secure 75-minute videoconference between Beijing and Vienna [317].

Building on these achievements, China launched a second quantum satellite, Jinan 1, aiming to conduct quantum key distribution experiments in lower-Earth orbit [318]. Jinan 1, significantly lighter than Micius, is designed to generate quantum keys at much higher speeds, which is crucial for real-time, satellite-to-ground quantum key distribution for a vast number of users globally. This marks a pivotal step towards establishing a robust, ultra-secure communications network with global coverage, showcasing significant improvements in satellite link efficiency and key generation rates necessary for a practical quantum communications network.

These are some progresses where quantum satellites overcome terrestrial limitations and enables global quantum communication. Such advancements suggest a future where quantum satellite constellations, combined with fiber quantum networks, could provide intercontinental connectivity, enhancing the quantum Internet's infrastructure. This approach would not only ensure ultra-secure communication across the globe but also support fundamental tests at the intersection of quantum physics and relativity, furthering our understanding of quantum phenomena.

However, the path to a fully operational Quantum Internet is fraught with challenges. Maintaining quantum coherence over long distances and time, scalability of quantum technologies, interoperability with existing networks, and robust quantum error correction methods are among the significant hurdles to overcome. The work in [319] focuses on the precision timing necessary for quantum networking, specifically improving timing resolution without new hardware through innovative designs. The research demonstrates significant reductions in accidental coincidences, enhancing the efficiency of quantum local area networks (QLANs). The integration of classical control planes for secure data management and the use of commercial quantum key distribution (QKD) systems for securing these planes are also highlighted, showcasing practical steps towards a secure quantum network infrastructure. The research in [320] addresses the challenge of coupling losses in on-chip quantum photonic systems, exploring fiber-to-chip couplers, including lateral couplers and vertical couplers. The study identifies the limitations and potentials of these couplers in reducing loss and enhancing the scale of quantum photonic systems. Innovations in generating narrowband entangled photon pairs and high-fidelity polarization-entangled states is discussed in [280]. This is achieved using silicon-on-insulator waveguides and III-V platform phase-matching techniques demonstrate significant progress in integrated quantum communications. These technologies are crucial for developing compact, efficient sources of quantum states for quantum networking. A review in [322] discusses recent advances in novel quantum materials for solid-state single-photon sources. The study highlights the development of sources that exhibit high efficiency, indistinguishability, and control over quantum state properties. These advancements are pivotal for quantum computing, communications, and sensing applications. Furthermore, research in quantum metaphotonics [323] focuses on increasing the spontaneous emission rate of quantum emitters (QEs) through the use of metal surfaces and dielectric nanostructures. This field has shown significant experimental progress in directing the generation of single photons with desired properties, such as specific polarizations and orbital angular momentum, through the design of meta-structures. Additionally, emerging applications and recent advances in quantum memories include techniques for rephasing collective atomic coherence and achieving storage in spin states of rare-earth-ion-doped ensembles [324]. These developments are crucial for the advancement of quantum computing and communication technologies, offering new possibilities for the storage and retrieval of quantum information.

Despite these challenges, the potential of quantum networks to transform secure communication, computational capabilities, and scientific research is immense. While the realization of a global Quantum Internet may still be at an experimental stage, ongoing research and development efforts are steadily paving the way for this quantum leap in networking technology, promising a future where information is transmitted, processed, and secured in ways previously thought impossible.

## 8 FUTURE RESEARCH AND OPEN PROBLEMS IN QUANTUM NETWORKS

Quantum networks leverage the principles of quantum mechanics and represent a frontier in the evolution of communication technologies. They promise to revolutionize our capabilities in secure communication, distributed quantum computing, and quantum information processing. However, despite significant progress, the path toward fully functional quantum networks is strewn with both technical and theoretical challenges. Addressing these challenges through future research is crucial for realizing the potential of quantum networks. Here, we explore some of the key open problems and areas for future research in this exciting field.

*Scalability and Integration:* One of the foremost challenges in quantum networks is scalability. Current quantum systems are relatively small-scale, limited by the difficulty in maintaining quantum coherence over large networks and the technological challenges of integrating quantum systems with existing infrastructure. Future research must focus on developing scalable quantum repeaters and error correction protocols that can maintain the fidelity of quantum states over long distances. Moreover, integrating quantum and classical networks requires innovative approaches to ensure seamless operation across fundamentally different technologies.

*Quantum Error Correction:* Quantum information is susceptible to errors from decoherence and operational imperfections, more so than classical information. Quantum error correction (QEC) is vital for preserving the integrity of quantum information over time and distance. Developing efficient and practical QEC codes that can be implemented in quantum networks remains a significant challenge. Research in this area not only supports quantum communication but is also crucial for the realization of quantum computing.

*Quantum Network Security:* While quantum key distribution (QKD) offers theoretically secure communication, practical implementations face vulnerabilities, including side-channel attacks, photon number splitting, timing attacks and other technological limitations. Also, the potential for quantum computing to break classical cryptographic schemes necessitates the development of quantum-resistant algorithms. Future research must address these vulnerabilities, developing more robust quantum cryptographic protocols and exploring quantum security beyond QKD, such as quantum secure direct communication (QSDC) and device-independent QKD.

*Entanglement Distribution and Management:* Entanglement is a resource that underpins the unique capabilities of quantum networks, including quantum teleportation and secure communication. Efficiently generating, distributing, and managing entanglement across a network, especially over long distances and through noisy channels, poses significant technical and theoretical challenges. Future research directions include the development of high-rate entanglement distribution protocols and the creation of quantum memories with long coherence times capable of storing entangled states with high fidelity. This includes the creation of multiplexed quantum communication channels to increase the throughput of quantum information.

Network Architectures and Protocols: The design of quantum network architectures and protocols is still in its infancy. Research is needed to develop protocols that can efficiently manage quantum and classical resources, optimize network routing and error correction, and support a variety of quantum network applications. Additionally, the exploration of quantum network topologies and their impact on performance and robustness is an open area of research.

Quantum Network Topologies and Traffic Management: Investigating optimal network topologies for quantum networks and developing traffic management protocols that can handle the unique requirements of quantum information are vital areas of research. This includes the study of how to best route quantum information through a network of nodes with varying capabilities and how to manage the network's classical and quantum data traffic efficiently.

Application-Specific Quantum Network Protocols: Beyond general-purpose quantum communication, tailoring quantum network protocols to specific applications such as distributed quantum computing, secure multi-party computation, and quantum sensing networks is an open area of research. Research into how quantum networks can enable new computational paradigms, enhance metrology, and support novel applications will be crucial for unlocking the full potential of quantum technologies.

Quantum Software and Simulation Tools: Developing software and simulation tools that can model quantum networks under realistic conditions is essential for the design and analysis of quantum communication systems. These tools need to account for quantum mechanical effects accurately, including entanglement, decoherence, and quantum noise, and provide insights into network performance and security.

Interoperability and Standards: As quantum networks move from theory to practice, establishing interoperability standards for quantum communication protocols and hardware is critical. This includes defining standards for quantum network interfaces, communication protocols, and security algorithms. Research in this area will facilitate the integration of quantum networks into existing infrastructure and ensure compatibility between different quantum systems and devices.

Interdisciplinary and Cross-Functional Research: Advancing quantum networks requires a multidisciplinary approach, merging insights from quantum physics, computer science, engineering, and information theory. Collaborative research efforts that bridge these disciplines are essential for addressing the complex challenges of quantum networks. Moreover, fostering partnerships between academia, industry, and government can accelerate the development of practical quantum network technologies and applications.

The future of quantum networks is ripe with both challenges and opportunities. Addressing the open problems in scalability, error correction, security, entanglement management, and network design will require concerted efforts across multiple disciplines. Through targeted research and development efforts, the advancements in these areas will pave the way for the realization of fully functional, secure, and efficient quantum networks, marking a significant leap forward in the field of quantum information science.